\documentclass[preprint,nofootinbib]{revtex4-1}
\usepackage{graphicx,epsfig,rotate}
\def\iso#1#2{\mbox{${}^{#2}{\rm #1}$}}
\def\he#1{\iso{He}{#1}}
\def\li#1{\iso{Li}{#1}}
\def\be#1{\iso{Be}{#1}}

\def\neff{N_{\rm eff}}

\def\avg#1{\langle #1 \rangle}

\def\ga{\mathrel{\raise.3ex\hbox{$>$\kern-.75em\lower1ex\hbox{$\sim$}}}}
\def\la{\mathrel{\raise.3ex\hbox{$<$\kern-.75em\lower1ex\hbox{$\sim$}}}}

\def\beq{\begin{equation}}
\def\eeq{\end{equation}}
\def\beqar{\begin{eqnarray}}
\def\eeqar{\end{eqnarray}}

\begin{document}

\rightline{UMN--TH--3432/15}
\rightline{FTPI--MINN--15/19}
\rightline{April 2015}

\title{Big Bang Nucleosynthesis: 2015}

\author{Richard H. Cyburt}
\affiliation{Joint Institute for Nuclear Astrophysics (JINA), National Superconducting Cyclotron Laboratory (NSCL), Michigan State University, East Lansing, MI 48824}
\author{Brian D. Fields}
\affiliation{Departments of Astronomy and of Physics, University of Illinois,
Urbana, IL 61801}
\author{Keith A. Olive}
\affiliation{William I. Fine Theoretical Physics Institute,
School of Physics and Astronomy,
University of Minnesota, Minneapolis, MN 55455, USA}
\author{Tsung-Han Yeh}
\affiliation{Departments of Astronomy and of Physics, University of Illinois,~
Urbana, IL 61801}

\begin{abstract}
Big-bang nucleosynthesis (BBN) describes the production of the
lightest nuclides via a dynamic
interplay among the four fundamental forces during
the first seconds of cosmic time.
We briefly overview the essentials of this physics,
and present new calculations of light element abundances
through \li6 and \li7, with updated nuclear reactions
and uncertainties including those in the neutron lifetime.
We provide fits to these results as a function of baryon density and of
the number of neutrino flavors, $N_\nu$.
We review recent developments in BBN,
particularly new, precision {\em Planck} cosmic microwave background (CMB)
measurements that now probe
the baryon density, helium content, and the effective number of degrees of freedom, $\neff$.
These measurements allow for a tight
test of BBN and of cosmology using CMB data alone. 
Our likelihood analysis convolves the 2015 {\em Planck} data chains with our BBN
output and observational data.
Adding astronomical measurements
of light elements strengthens the power of BBN. We include a new determination of 
the primordial helium abundance in our likelihood analysis. 
New D/H observations are now more precise than the
corresponding theoretical predictions, and are consistent with
the Standard Model and the {\em Planck} baryon density. Moreover,
D/H now provides a tight measurement of $N_\nu$ when combined with the CMB
baryon density, and provides a $2\sigma$ upper limit $N_\nu < 3.2$.
The new precision of the CMB and of D/H observations
together leave D/H predictions as the largest source of uncertainties.
Future improvement in BBN calculations will therefore rely on improved 
nuclear cross section data.  
In contrast with D/H and \he4, \li7 predictions continue to disagree with observations,
perhaps pointing to new physics.
We conclude with a look at future directions
including key nuclear reactions, astronomical observations,
and theoretical issues.
\end{abstract}

\pacs{}
\keywords{}

\maketitle

\section{Introduction}

Big bang nucleosynthesis (BBN) is one of few probes of the very early
Universe with direct experimental or observational consequences
\cite{bbn,bbn2,iocco}.  In the context of the Standard Models of cosmology
and of nuclear and particle physics, BBN is an effectively parameter-free theory \cite{cfo2}.
Namely, standard BBN (SBBN)
assumes  spacetime characterized by General Relativity and the
$\Lambda$CDM cosmology, and  microphysics
characterized by Standard Model particle content and interactions,
with three light neutrino species, with negligible
effects due to dark matter and dark energy.

In SBBN,  the abundances of the four light nuclei are usually parameterized by
the baryon-to-photon ratio $\eta \equiv n_b/n_\gamma$, or
equivalently the present
baryon density, $\Omega_b h^2 \equiv \omega_b$.
This quantity has been fixed by a series of precise measurements of
microwave background anisotropies, most recently by {\it Planck} yielding
$\eta = 6.10 \pm 0.04$ \cite{Planck2015}.
Thus the success or failure of SBBN rests solely
on the comparison of theoretical predictions with observational determinations.

While precise predictions from SBBN are feasible, they rely on
well-measured cross sections and a well-measured neutron lifetime.
Indeed, even prior to the {\em WMAP} era,
theoretical predictions for D, \he3, and \he4 were reasonably accurate,
however, uncertainties in nuclear cross sections leading to \be7 and 
\li7 were relatively
large. Many modern analyses of nuclear rates for BBN were based on the NACRE
compilation \cite{nacre} and recent BBN calculations by several groups
are in good agreement \cite{bn,coc0,cfo1,cvcr01, cfo3,coc,cyburt, coc2,cuoco,coc3}.
Recent remeasurements of the \he3($\alpha,\gamma$)\be7 cross section \cite{he3alpha}
did improve the theoretical accuracy of the prediction but exacerbated the discrepancy
between theory and observation \cite{cfo5}. Very recently, the NACRE collaboration
has issued an update (NACRE-II) of its nuclear rate tabulation \cite{nacreII}. These were first used
in \cite{coc4} and we incorporate the new rates in the results discussed below.

The neutron mean life has had a rather sordid history. Because it scales the weak interaction
rates between $n \leftrightarrow p$, the neutron mean life controls the neutron-to-proton
ratio at freeze-out and directly affects the \he4 abundance (and the other light elements to a
lesser extent). The value $\tau_n = 918 \pm 14 \ \rm s$
reported by Cristensen et al. in 1972 \cite{christ} dominated
the weighted mean for the accepted value through mid 1980s. Despite the low value of 877 $\pm$ 8 s
reported by
Bondarenko et al. in 1978 \cite{bond}, the `accepted' mean value (as reported in the
``Review of Particle Physics'') remained high, and the high value was reinforced by
a measurement by Byrne et al. \cite{byrne} in 1980 of 937 $\pm$ 18 s.  The range 877 s - 937 s
was used by Olive et al. \cite{ossty} to explore the sensitivity of BBN predictions to this
apparently uncertain quantity treated then as an uncertain input parameter to BBN
calculations (along with the number of light neutrino flavors $N_\nu$ and the
baryon-to-photon ration, $\eta$). In the late 1980's a number of lower measurements began to surface,
and in 1989 Mampe et al. \cite{mampe} claimed to measure a mean life of 877.6 $\pm$ 3 s (remarkably
consistent with the current world average). Subsequently, it appeared that questions regarding the
neutron mean life had been resolved, as the mean value varied very little between 1990 and 2010,
settling at 885.6 $\pm$ 0.8. However, in 2005, there was already a sign that the mean life was about to
shift to lower values once again.  Serebrov et al. \cite{sere} reported a very precise measurement of
878.5 $\pm 0.8$ which was used in BBN calculations in \cite{mks}. This was followed by several more recent measurements and reanalyses
tending to lower values so that the current world average is \cite{rpp} $\tau_n = 880.3 \pm 1.1$.
We will explore the impact of the new lifetime on the light element abundances.

On the observational side,
the \he4 abundance determination comes most directly from 
measurements of emission lines in
highly ionized gas in nearby low-metallicity dwarf galaxies
(extragalactic HII regions).
The helium abundance uncertainties are dominated by systematic effects \cite{osk,aos,aos3}.
The model used to determine the \he4 abundance contains eight physical parameters,
including the \he4 abundance, that is used to predict a set of ten H and He emission line
ratios, which can be compared with observations \cite{it,its07,isg13,itg}. 
Unfortunately, there are only a dozen or
so observations for which the data and/or model is reliable, and even in those cases,
degeneracies among the parameters often lead to relatively large uncertainties for each system as
well as a large uncertainty in the regression to zero metallicity.
Newly calculated \he4 emissivities \cite{pfsd}  and the addition of a new near infra-red line \cite{itg} have led to
lower abundance determinations \cite{aops,aos4}, bringing the central value of the \he4 abundance
determination into good agreement with the SBBN prediction.
Moreover, {\em Planck} measurements of CMB anisotropies are now
precise enough to give interesting measures of primordial helium
via its effect on the anisotropy damping tail.

New observations and analyses of
quasar absorption systems have dramatically improved the observational
determination of D/H.  Using a a handful of systems where accurate determinations can be made,
Cooke et al. \cite{cooke} not only significantly lowered the uncertainty in the mean D/H abundance, but the dispersion present in old data has all but been erased. Because of its sensitivity to the baryon density,
D/H is a powerful probe of SBBN and now the small uncertainties in both the data
and prediction become an excellent test of concordance between SBBN and the CMB determination
of the baryon density.

In contrast to the predicted abundances of \he4 and D/H, the \li7 abundance shows a definite
discrepancy with all observational determinations from halo dwarf stars. To date, there is no solution
that is either not tuned or requires substantial departures from Standard Model physics.
Attempts at solutions include modifications of the nuclear rates \cite{coc3,cfo4,boyd}
or the inclusion of new resonant interactions \cite{cp,chfo,brog}; stellar depletion \cite{dep};
lithium diffusion in the post-recombination universe \cite{diff};
new (non-standard model) particles decaying around the time of BBN \cite{jed,decay,ceflos1.5,serp}; axion
cooling \cite{sik}; or variations in the fundamental constants \cite{dfw,cnouv}.

In this review, we survey the current status of SBBN theory and its compatibility with
observation. Using an up-to-date nuclear network, we present new Monte-Carlo
estimates of the theoretical predictions based on a full set of nuclear cross sections and
their uncertainties. We will highlight those reactions that still carry the greatest uncertainties
and how those rates affect the light element abundances. 
We will also highlight the effect of the new determination of the neutron mean life on the \he4
abundance and the abundances of the other light elements as well.

Compatibility with observation is demonstrated by the construction of likelihood functions for
each of the light elements \cite{cfo1} by convolving individual  theoretical and observational
likelihood functions.  
Our BBN calculations are also convolved
with data chains provided by the 2015 {\em Planck} data release \cite{Planck2015}. 
This allows us to construct 2-dimensional ($\eta, Y_p$) and 3-dimensional ($\eta, Y_p, N_\nu$) 
likelihood distributions.
Such an analysis is timely and important given the recent
advances in the \he4 and D/H observational landscape. In fact,
despite the accuracy of the SBBN prediction for D/H, the tiny uncertainty in
observed D/H  now leads to a likelihood dominated by theory errors.
The tight agreement between D/H prediction and observation is in sharp
contrast to the discrepancy in \li7/H.  We briefly discuss this ``lithium problem,''
and discuss recent nuclear measurements that rule out a nuclear fix to this problem,
leaving as explanations either astrophysical systematics or new physics.
While the \li6 was shown to be an artifact of astrophysical systematics \cite{cayrel,lind}, for now,
the \li7 problem persists.

As a tool for modern cosmology and astro-particle physics, SBBN is a powerful
probe for constraining physics beyond the Standard Model \cite{cfos}.  Often, these
constraints can be parameterized by the effect of new physics on the speed-up of
the expansion rate of the Universe and subsequently translated into a limit
on the number of equivalent neutrino flavors or effective number of relativistic 
degrees of freedom, $\neff$. We update
these constraints and compare them to recent limits on $\neff$
from the microwave background anisotropy and large scale structure.
We will show that for the first time, D/H provides a more stringent constraint
on $\neff$ than the \he4 mass fraction.

The structure of this review is as follows:
In the next section, we discuss the relevant updates to the nuclear rates used in the
calculation of the light element abundances. We also discuss the sensitivity to the neutron mean life.
In this section, we present our base-line results for SBBN assuming the {\em Planck} value for $\eta$,
the RPP value for $\tau_n$ and $N_\nu = 3$. In section III, we briefly review the values of the
observational abundance determinations and their uncertainties that we adopt for comparison with the SBBN calculations
from the previous section. In section IV, we discuss our Monte Carlo methods, and
construct likelihood functions for each of the light elements, and we extend these methods
to discuss limits on $N_\nu$ in section V. The lithium problem is summarized
in section VI. 
A discussion of these results and the future outlook appears in section VII.

\section {Preliminaries}

\subsection{SBBN}

As defined in the previous section,
SBBN refers to BBN in the context of Einstein gravity, with a Friedmann-Lema\^{i}tre-Robertson-Walker
cosmological background. It assumes the Standard Model of nuclear and particle physics, or in other words,
the standard set of nuclear and particle interactions and nuclear and particle content.
Implicitly, this means a theory with a number $N_\nu = 3$ of very light neutrino flavors\footnote{We
distinguish between the number of neutrino flavors, three in the Standard Model, and $\neff$, equal to 3.046 in the Standard Model, which corresponds to the effective number of neutrinos present in the thermal
bath due to the higher temperature from $e^+ e^-$ annihilations before neutrinos are completely decoupled.}.
SBBN also makes the well-justified assumption 
that during the epoch of nucleosynthesis, the universe was radiation dominated,
so that the dominant component of the energy density of the universe can be expressed as
\begin{equation}
\rho = {\pi^2 \over 30} \left( 2 + {7 \over 2} + {7 \over 4}N_\nu \right) T^4 ,
\label{rho}
\end{equation}
taking into account the contributions of photons, electrons and positrons, and 
neutrino flavors appropriate for temperatures $T > 1$ MeV. At these temperatures, weak
interaction rates between neutrons and protons maintain equilibrium.

At lower temperatures, the weak interactions can no longer keep up with the expansion
of the universe or  equivalently,
the mean time for an interaction becomes longer than the age of the Universe.
Thus, the freeze-out condition is set by
\beq
G_F^2 T^5 \sim \Gamma_{\rm wk} (T_f)  = H(T_f) \sim G_N^{1/2} T^2,
\label{fo}
\eeq
where $\Gamma_{\rm wk}$ represents the relevant weak interaction rates
per baryon
that scale roughly as $T^5$, and $H$ is the Hubble parameter 
\beq
H^2 = \frac{8\pi}{3} G_N \rho
\label{hubb}
\eeq
and scales as $T^2$
in a radiation dominated universe. $G_{F}$ and $G_N$ are the Fermi and Newton constants respectively.
Freeze-out occurs when the weak interaction rate 
falls below the expansion rate,
$\Gamma_{wk} < H$. The $\beta$-interactions that control
the relative abundances of neutrons and protons freeze out at $T_f \sim \rm 0.8 MeV$.
At freeze-out, the neutron-to-proton ratio is given
approximately by the Boltzmann factor,
$(n/p)_f \simeq e^{-\Delta m/T_f} \sim 1/5$, where $\Delta m=m_n-m_p$ is the neutron--proton
mass difference. After freeze-out, free neutron decays drop the ratio
slightly to $(n/p)_{\rm bbn} \simeq 1/7$ before nucleosynthesis begins.
A useful semi-analytic description
of freeze-out can be found in~\cite{bbf,muk}.

The first link in the nucleosynthetic chain is  $p+n \rightarrow$ d $+~\gamma$
and although the binding energy of deuterium is relatively small, $E_B = 2.2$ MeV,
the large number of photons relative to nucleons,
$\eta^{-1} \sim 10^{9}$ causes the so-called deuterium bottleneck.
BBN is delayed until $\eta^{-1} {\rm exp}(-E_B/T) \sim 1$ when the
deuterium destruction rate finally falls below its production rate.
This occurs when the temperature is approximately 
$T \sim E_B/\ln \eta^{-1} \sim 0.1 \ \rm MeV$.

To a good approximation, almost all of the neutrons present when
the deuterium bottleneck breaks end up in \he4. It is therefore very easy
to estimate the \he4 mass fraction,
\begin{equation}
Y_p = \frac{2(n/p)}{1 + (n/p)} \approx 0.25 ,
\label{ynp}
\end{equation}
where we have evaluated $Y_p$ using $(n/p) \approx 1/7$.
The other light
elements are produced in significantly smaller abundances, justifying
our approximation for the \he4 mass fraction.  D and \he3 are produced at the level of about $10^{-5}$ by
number,  and \li7 at the level of $10^{-10}$ by number.
A cooling Universe, Coulomb barriers, and the mass gap at
$A=8$ prevents the production of other isotopes in any
significant quantity.

\subsection{Updated Nuclear Rates}

Our BBN results use an updated version of our code \cite{cfo1,cfo5},
itself a descendant of the Wagoner code \cite{wagoner}.  For the weak
$n\leftrightarrow p$ interconversion rates, the code calculates the 1-D
phase space integrals at tree level; this corresponds to the
assumption that the nucleon remains at rest.  The weak
$n\leftrightarrow p$ interconversion rates are normalized such that we
recover the adopted mean neutron lifetime at low temperature and
density.  To this we added order-$\alpha$ radiative and bremsstrahlung
QED corrections, and included Coulomb corrections for reactions with
$p e^-$ in the initial or final states \cite{dicus}.  We neglected
additional corrections, because the overall contribution of all other
effects is relatively insignificant.  Finite temperature radiative
corrections leads to $\Delta Y_p \sim$ 0.0004; corrections to electron
mass leads to $\Delta Y_p \sim$ -0.0001; neutrino heating due to
$e^{+}e^{-}$ annihilation leads to $\Delta Y_p \sim$ 0.0002
\cite{dicus}.  Problems in the original code due to the choice of time-steps
in the numerical integration \cite{kernan} have been corrected here. 
Finally, we included the effects of finite nucleon mass
\cite{seckel} by increasing the final \he4 abundance with $\Delta Y_p$
= + 0.0012.  We formally adopt the Particle Data Group's current recommended
value~\cite{rpp}: $\tau_n = 880.3 \pm 1.1$ for the mean free neutron
decay lifetime, and assume it is normally distributed.  

In addition to the weak $n \leftrightarrow p$ interconversion rates,
BBN relies on well-measured cross sections.  The latest update to
these reaction rates has been evaluated by the NACRE collaboration and
released as NACRE-II~\cite{nacreII}.  Only charge-induced reactions
were considered in NACRE-II, and many more reactions are evaluated
there than is relevant for BBN.  Those reactions of relevance
are shown in table~\ref{tab:nacre2}.  The compilation presents tables
of reaction rates, with recommended ``adopted'', ``low'' and ``high''
values.  We assume the rates are distributed with a log-normal
distribution with: \beqar
\mu&\equiv&\ln{\sqrt{R_{high}\!\times\!R_{low}}}\\
\sigma&\equiv&\ln{\sqrt{R_{high}/R_{low}}}, \eeqar where $R_{low}$ and
$R_{high}$ are the recommended ``low'' and ``high'' values for the
reaction rates, respectively~\cite{coc4}.  These rates agree well with previous
evaluations~\cite{cfo1,cyburt,coc2,cuoco}, largely because they depend on the same experimental
data.  Uncertainties in
the NACRE-II evaluation are similar to, but tend to be larger than
previous studies.  This may stem from the method used to
derive accurate descriptions of the data and the fitting potential models
used in Distorted-Wave Born Approximation (DWBA) calculations.

\begin{table}[ht]
\caption{Reactions of relevance for BBN from the NACRE-II compilation.
\label{tab:nacre2}
}
\begin{tabular}{|c|c|c|}
\hline
 $d(p,\gamma )\he3$ &
 $d(d,\gamma )\he4$ &
 $d(d,n)\he3$ \\
\hline
 $d(d,p)t$ &
 $t(d,n)\he4$ &
 $\he3(d,p)\he4$ \\
\hline
 $d(\alpha ,\gamma )\li6$ &
 $\li6(p,\gamma )\be7$ &
 $\li6(p,\alpha )\he3$ \\
\hline
 $\li7(p,\alpha )\he4$ &
 $\li7(p,\gamma )\be8$\footnote{\be8 is not in our nuclear network, \be8 is assumed to spontaneously decay into 2\he4's.} & \\
\hline
\end{tabular}
\end{table}

The remaining relevant $n$-induced rates, $p(n,\gamma )d$,
$\he3(n,p)t$ and $\be7(n,p)\li7$, need to be taken from different
sources.  We adopt the evaluation from~\cite{ando} for the key
reaction $p(n,\gamma )d$.  The remaining $(n,p)$ reactions we use
are rates taken from~\cite{cyburt}.  We choose log-normal parameters in such a
way to keep the means and variances of the reaction rates invariant.

\subsection{First Results}
As a prelude to the more detailed analysis given below, 
we first discuss the BBN predictions at a fixed value of $\eta$.
This benchmark can then be compared to the results of other codes.

Here, we fix $\eta_{10} \equiv 10^{10} \eta = 6.10$.  This is related to the value of $\omega_b$ determined
by {\em Planck} \cite{Planck2015} based on a combination of  temperature and polarization
data\footnote{A straight interpretation of the {\em Planck} result based on the TT,TE,EE,+lowP anisotropy data would yield $\eta_{10} = 6.09$.  However
this result already includes the He abundance from BBN. Our choice of $\eta$ 
is discussed in Section IV.C and Table IV below.}. The result of our BBN calculation at $\eta_{10} = 6.10$
can be found in Table
\ref{tab:re} compared to the fit in \cite{iocco},
based on the PArthENoPE code \cite{pisanti}. 
As one can see, the results of the two 
codes are in excellent agreement for all of the light elements. The results can be quickly compared
with the observed abundances given in the next section.  However a more rigorous treatment
of the comparison between theory and observation will be given in section IV. 

\bigskip

\begin{table}[ht]
\caption{Comparison of BBN Results
\label{tab:re}
}
\begin{tabular}{|l|c|c|c|c|c|c|c|}
\hline
 & $\eta_{10}$ & $N_{\nu}$ & $Y_p$ & D/H & \he3/H & \li7/H & \li6/H \\
\hline
This Work& 6.10 & 3 & 0.2470 & 2.579 $\times$ $10^{-5}$& 0.9996 $\times$ $10^{-5}$ &  $4.648 \times$ $10^{-10}$ & 1.288 $\times$ $10^{-14}$ \\
\hline
Iocco et al.~\cite{iocco} fit & 6.10 & 3 & 0.2463 & 2.578 $\times$ $10^{-5}$& 0.9983 $\times$ $10^{-5}$ & 4.646 $\times$ $10^{-10}$ & 1.290 $\times$ $10^{-14}$ \\
\hline
\end{tabular}
\end{table}

\section{Observations}
\label{sect:obs}

Before making a direct comparison of the SBBN results, we
first discuss astrophysical observations of
light elements. Here we focus on \he4, D, and the Li isotopes,
all of which are accessible in primitive environments making it
possible to extrapolate existing observations to their primordial abundances.
BBN also produces \he3 in observable amounts,
and \he3 is detectable via its hyperfine emission line,
but this line is only accessible within Milky Way gas clouds
that are far from pristine
\cite{he3obs}. Because of the uncertain post-BBN nucleosynthetic history of \he3,
it is not possible to use these high-metallicity environments
to infer primordial \he3 at a level useful for probing BBN
\cite{he3cos}.
After an overview of the observational status of the four remaining isotopes,
we  turn to the CMB, which includes
constraints not only on the baryon density
but now also on \he4 and $\neff$.

\subsection{Helium-4}

\he4 has long since been the element of choice for setting constraints on physics
beyond the Standard Model.  The reasoning is simple:  as discussed above, the \he4 abundance is
almost completely controlled by the number of free neutrons at the onset of
nucleosynthesis, and that number is determined by the freeze-out of the weak
$n \leftrightarrow p$ rates. The resulting mass fraction of \he4 is given in Eq. (\ref{ynp}).
As we have seen, the SBBN result for the $Y_p$ dependence on the baryon density
is only logarithmic and therefore
\he4 is not a particularly good baryometer. Nevertheless,
 it is quite sensitive to any changes in the freeze-out temperature, $T_f$, through the relation
(\ref{fo}). However, strong limits on
physics beyond the Standard Model \cite{cfos} requires accurate \he4 abundances from observations.

The \he4 abundance is determined by measurements of He (and H) emission lines in
extragalactic HII regions. Since \he4 is produced in stars along with heavier elements,  the primordial mass fraction of \he4, $Y_{p} \equiv \rho(\he4)/\rho_b$, is determined by a regression of the  helium abundance versus metallicity \citep{ptp74}.   However, due to numerous systematic uncertainties, obtaining an accuracy better than 1\%  in the primordial helium abundance is very difficult \citep{osk,its07,plp07}. The theoretical model
that is used to extract a \he4 abundance contains eight physical parameters to predict the fluxes of
nine emission line ratios that can be compared directly with observations \footnote{Below, we will use results based on the inclusion of a tenth line (seventh He line) seen in the near infra-red \cite{itg}.}.
 The parameters include, the electron density, optical depth, temperature, equivalent widths of underlying
 absorption for both H and He, a correction for reddening, the neutral hydrogen fraction, and of course
 the \he4 abundance. Using theoretical emissivities, the model can be used to predict the fluxes
 of 6 He lines (relative to H$\beta$) as well as 3 H lines (also relative to H$\beta$).
 The lines are chosen for their ability to break degeneracies among the inputs when possible.
 For a recent discussion see \cite{aops,aos4}.

There is a considerable amount of \he4 data available \cite{it,its07}. A Markov Chain Monte Carlo
analysis of the 8-dimensional parameter space for 93 H~II regions reported in \cite{its07} was performed in
\cite{aos3}.  By marginalizing over the other 7 parameters, the \he4 abundance (and its uncertainty)
can be determined. However, for most the data, the $\chi^2$s obtained by comparing
the theoretically derived fluxes for the 9 emission lines with those observed, were typically very large
($\gg 1$) indicating either a problem with the data, a problem with the model, or problems with both.
Selecting only data for which 6 He lines were available, and a $\chi^2 < 4$, left only 25 objects for
the subsequent analysis. Further cuts of solutions with for example,  anomalously high neutral H fractions, or
excessive corrections due to underlying absorptions brought the sample down to 14 objects
that yielded $Y = 0.2534 \pm 0.0083 + (54 \pm 102)$O/H based on a linear regression and
$Y_p = 0.2574 \pm 0.0036$ based on a weighted mean of the same data. 

Recently, a new analysis of the theoretical emissivities has been performed \cite{pfsd}. This includes improved
photoionization cross-sections and a correction of errors found in the previous result.
The new emissivities are systematically higher, and for some lines the increase in the emissivity is
3-5\% or higher. As a consequence, one expects lower \he4 abundances
using the new emissivities.   In \cite{aops}, the same initial data were used with the same
quality cuts, now leaving 16 objects in the final sample. Individual objects  typically showed
5-10 \% lower \he4 abundance yielding $Y = 0.2465 \pm 0.0097 + (96 \pm 122) {\rm O/H}$ for a regression.
Once again, one could argue that the lack of true indication
of a slope in the data over the restricted baseline may justify using the mean rather than the regression.
The mean was then found to be $Y_p = 0.2535 \pm 0.0036$. 
The large errors in $Y_p$ determined from the regression were 
 due to a combination of large errors on individual objects,
a relatively low number of objects with $\chi^2 < 4$, a short baseline in O/H, and a poorly
determined slope (though the analysis using the new emissivities does show more positive evidence for a slope of Y vs O/H). 

More recently, new observations include a near infrared
line at $\lambda 10830$ \cite{itg}. The importance of this line stems from
its dependence on density and temperature that differs from other observed He lines.
This potentially breaks the degeneracy seen between these two parameters
that is one of the major culprits for large  uncertainties in \he4 abundance determinations.
There are 16 objects satisfying $\chi^2 \la 6$ \cite{aos4} (there are now two degrees for freedom
rather than one), with all seven He measured (though these are not exactly the same 16 objects used in 
\cite{aops}).
Indeed it was found \cite{aos4} that the inclusion of this line, did in fact reduce
the uncertainty and leads to a better defined regression 
\beq
Y_{p} = 0.2449 \pm 0.0040 + (78.9 \pm 43.3) {\rm O/H}
\label{ypr}
\eeq
Unlike past analyses, there is now a well-defined slope in  the regression, 
making the mean,
$Y_p = 0.2515 \pm 0.0017$ less justifiable as an estimate of primordial \he4. 
The benefit of adding the IR He line is seen to reduce the uncertainty in $Y_p$
by over a factor of two. This is due to the better determined abundances of individual objects
 and a better determined slope. It remains the case, however, that most of the 
available observational data are not well fit by the model. 
Comparing with the value of $Y_p$ given in Table \ref{tab:re}, 
the intercept of the regression (\ref{ypr}) is in good agreement with the results of SBBN.

\subsection{Deuterium}
\label{sect:deut}

Because of its strong dependence on the baryon density, deuterium is an excellent
baryometer.  Furthermore, since there are no known astrophysical sources for
deuterium production \cite{els} and thus all deuterium must be of primordial origin, any
observed deuterium provides us with an upper bound on the baryon-to-photon ratio
\cite{rafs,ggst}.
However, the monotonic decrease in the deuterium abundance over time indicates that
the galactic chemical evolution \cite{gce,opvs}
affects the interpretation of any local measurements of the deuterium abundance
such as in the local interstellar medium \cite{lism}, galactic disk \cite{gd}, or galactic halo \cite{gh}.

The role of D/H in BBN was significantly promoted when measurements of
D/H ratios in quasar absorption systems at high redshift became available.
In a short note in 1976, Adams \cite{adams} outlined the conditions that would permit
the detectability of deuterium in such systems.  However it was not until 1997, that the
first reliable measurements of D/H at high redshift became available \cite{reliable}
(we do not discuss here the tumultuous period with conflicting high and low measures of D/H).
Over the next 20+ years, only a handful of new observations became available with abundances in
the range $\rm D/H = (1 - 4) \times 10^{-5}$ \cite{reliable,otherD}.  Despite the fact that there was considerable
dispersion in the data (unexpected if these observations correspond to primordial D/H),
the weighted mean of the data gave D/H = $(3.01 \pm 0.21)  \times 10^{-5}$ with a sample variance of 0.68.
While the data were in reasonably good agreement with the SBBN predicted value (discussed in detail below)
using the CMB determined value for the baryon-to-photon ratio, the dispersion indicated that either
the quoted error bars were underestimated and larger systematic errors were unaccounted for,
or if the dispersion was real, in situ destruction of deuterium must have taken place within these absorbers.
In the latter case, the highest ratio ($\sim 4\times 10^{-5}$) should be taken as the post-BBN value,
leaving room for some post-BBN production of D/H that may have accompanied the destruction of \li7
- we return to this possibility (or lack thereof) below.

In 2012, Pettini and Cooke \cite{pc} published results from a new observation of an absorber at $z = 3.05$,
with $\rm D/H = (2.54 \pm 0.05) \times 10^{-5}$ corresponding to an uncertainty of about 2\%
that can
be compared with typical uncertainties of 10 -- 20 \% in previous observations. This was followed
another precision observation and a reanalysis of the 2012 data along with a reanalysis of
a selection of three other objects from the literature (chosen using a strict set of restrictions
to be able to argue for the desired accuracy)~\cite{cooke}. The resulting set of five absorbers yielded
\beq
\left( \frac{{\rm D}}{\rm H} \right)_p = \left( 2.53 \pm 0.04 \right) \times 10^{-5}
\label{d/h}
\eeq
with a sample variance of only 0.05.
We will use this value in our SBBN analysis below\footnote{Note that the most recent measurement
described in \cite{cars} has a somewhat larger uncertainty, and its inclusion does not 
affect the weighted mean in Eq. (\ref{d/h})}.

\subsection{Lithium}

\label{sect:LiObs}

Lithium has by far the smallest observable primordial abundance
in SBBN, but as we will see provides an important consistency check
on the theory -- a check that currently is not satisfied!
In SBBN, mass-7 is made in the form of stable \li7, but also as 
radioactive \be7.
In its neutral form, \be7 decays via electron capture
with a half-life of 53 days.
In the early universe, however, the decay is delayed until the universe is cool enough
that \be7 can finally capture an electron at $z \sim 30,000$~\cite{ks}, shortly before
hydrogen recombination!
Thus \be7 decays long after the $\sim 3$ min timescale of BBN,
yet after recombination, all
mass-7 takes the form of \li7. Consequently,
\li7/H theory predictions  sum both mass-7 isotopes.
Note also that \li7 production
dominates at low $\eta$, while \be7 dominates at high $\eta$,
leading to
the characteristic ``lithium dip'' versus baryon density
in the Schramm plot (Fig.~\ref{fig:schramm}) described below.

A wide variety of astrophysical processes have been proposed as lithium
nucleosynthesis sites operating after BBN.
Cosmic-ray interactions with diffuse interstellar (or intergalactic) gas
produces both \li7 and \li6 via spallation
reaction such as $p_{\rm cr} + \iso{O}{16}_{\rm ism} \rightarrow \li{6,7} + \cdots$, and fusion $\he4_{\rm cr} + \he4_{\rm ism} \rightarrow \li{6,7} + \cdots$\cite{crnuke}.
In the supernova ``$\nu$-process,''
neutrino spallation reactions can also produce \li7
in the helium shell via $\nu + \he4 \rightarrow \he3$ followed by
$\he3+\he4 \rightarrow \be7 + \gamma$ as well as the mirror version of these
\cite{nupro} though the importance of this contribution to \li7 is limited by associated $^{11}$B production \cite{vosp}.
Finally, in somewhat lower mass stars undergoing the late,
asymptotic giant branch phase of evolution,
\he3 burning leads to high surface Li abundances,
some of that may (or may not!) survive to be ejected in the death of the stars
\cite{cf}.
Thus, despite its low abundance, \li7 is the only element
with significant production in the Big Bang, stars, and cosmic rays;
by contrast, the only conventional site of \li6 production is in cosmic
ray interactions \cite{rafs,li6nuke}.

To disentangle the diverse Li production processes observationally thus requires
measurements of lithium abundances as a function of metallicity.
As with \he4, the lowest-metallicity data should have negligible
Galactic contribution and point to the primordial abundances.
To date, the only systems for which such a metallicity evolution
can be traced are in metal-poor (Population II)
halo stars in our own Galaxy.  As shown by
Spite \& Spite \cite{spites},
halo main sequence (dwarf/subgiant) stars with
temperatures $T_{\rm eff} \ga 6000$ K have a constant
Li abundance, while Li/H decreases markedly for cooler stars.
The hotter stars have thin convection zones and so Li is not
brought to regions hot enough to destroy it.  These hot halo stars
that seem to preserve their Li are thus of great cosmological interest.
Spite \& Spite \cite{spites} found that these
stars with $[\rm Fe/H] \equiv \log[(Fe/H)/(Fe/H)_\odot] \la -1.5$ have a substantially lower Li content than solar metallicity stars,
and moreover the Li abundance does not vary with metallicity.
This ``Spite plateau'' points to the primordial origin of Li.
Furthermore, the Li/H abundance at the plateau
gives the primordial value {\em if the host stars have not
destroyed any of their lithium.}

The 1982 Spite plateau discovery was based on 11 stars
in the plateau region. Since then, the number of stars on the plateau
have increased by more than an order of magnitude.
Increasingly precise observations showed that the
scatter in Li abundances is very small for halo stars with
metallicities down to $[{\rm Fe/H}] \sim -3$ \cite{rnb,ryan2000,melendez2004,bonifacio2007,hos}.

Recently, thanks to huge increases in the numbers of known
metal-poor halo stars, Li data has ben extended to very low metallicity.
Surprisingly, at metallicities below about
$[{\rm Fe/H}] \la -3$, the Li/H scatter becomes large, in contrast
to the small scatter in the Spite plateau found at slightly higher
metallicity.
In particular, the trend in extremely metal-poor stars
is that
no stars have Li/H above the Spite plateau value,
a few are found near the plateau, but many lie significantly below
\cite{aoki2009,sbordone2010}.
This ``meltdown'' of the Spite plateau remains difficult to
understand from the point of view of stellar evolution, but
in any case seems to demand that at least {\em some} halo stars
have destroyed their Li.
Moreover, as we will see below,
this possibility has important consequences for the primordial lithium problem.

While the low-metallicity Li/H behavior is not understood,
the Spite plateau remains at $[{\rm Fe/H}] \simeq -3$ to $-1.5$;
lacking a clear reason to discard these data, we will use them
as a measure of primordial Li.  Following ref.~\cite{sbordone2010}
we adopt their average of the non-meltdown halo stars
 having
$[{\rm Fe/H} ] \ga -3$, giving
\beq
\label{eq:Li_p}
\left( \frac{{\rm Li}}{{\rm H}} \right)_{p} = (1.6  \pm 0.3) \times 10^{-10} \, .
\eeq
The stars in this sample were
observed and analyzed in a uniform way,
with Li abundances having been inferred from
the absorption line spectra using sophisticated 3D stellar
atmosphere models
that do not assume local thermodynamic equilibrium (LTE).

Most halo star observations measure only elemental Li,
because thermal broadening in the stellar atmospheres
exceeds the isotope separation between \li7 and \li6.
However, very high signal-to-noise measurements
are sensitive to asymmetries in the $\lambda6707$ lineshape that encode
this isotope information.  There were recent claims of \li6 detections,
with isotope ratios as high as $\li6/\li7 \la 0.1$ \cite{asplund}.
The implied \li6/H abundance lies far above the SBBN value,
leading to a putative ``\li6 problem.''

However, recent analyses with
3D, non-LTE stellar atmosphere models included
surface convection effects (akin to solar granulation),
and showed that these can entirely explain the observed
line asymmetry \cite{cayrel,lind}.  Thus case for \li6 detection in halo stars
and for a \li6 problem, is weakened.
Rather, the highest claimed \li6/\li7 ratio is
best interpreted as an upper limit.

\subsection{The Cosmic Microwave Background}

The cosmic microwave background (CMB) provides us with a snapshot of the 
universe at recombination ($z_{\rm rec} \simeq 1100$), and
encodes a wealth of cosmological information at unprecedented
precision.  
In particular, the CMB provides a particularly robust,
precise
measure of the cosmic baryon content, in a manner
completely independent of BBN \cite{wmap1,wmap,wmap2,planck,Planck2015}.  Recently, the CMB 
determinations of $\neff$ and $Y_p$ have also become quite interesting.
In this section we will see
how the precision of the CMB has changed the role
of BBN in cosmology, and enhanced the leverage of BBN to
probe new physics.

The physics of the CMB, and its relation to cosmological
parameters, is recounted in excellent reviews such as
\cite{cmbrev}.  
Here, we briefly and qualitatively summarize some of the key 
physics of the recombination epoch ($z_{\rm rec} \simeq 1100$) when the CMB was released.

Tiny primordial density fluctuations are laid down in the very early
universe, by inflation or some other mechanism.
After matter-radiation equality, $z_{\rm eq} \simeq 3400$, the dark matter density fluctuations
grow and form increasingly deep gravitational potentials.
The baryon-electron plasma is attracted to the potential wells,
and undergoes adiabatic compression as it falls in.
Prior to recombination, the plasma remains tightly coupled
to the CMB photons, whose pressure
$P_\gamma \propto T^4$ acting as a restoring force,
with sound speed $c_s^2 \sim 1/3$.  
This interplay of forces leads to
acoustic oscillations of the plasma.
The oscillations continue until (re)combination, 
when the universe goes from an opaque plasma to a transparent gas
of neutral H and He.  The decoupled photons for the most part
travel freely thereafter until detection today, recording a 
snapshot of the recombining universe.

Because the density fluctuations are small, perturbations
in the cosmic fluids are very well described by linear theory,
wherein different wavenumbers evolve independently.
Modes that have just attained their first contraction to a density maximum
at recombination
occur at the sound horizon $\sim c_s t_{\rm rec}$.
This lengthscale projects onto the sky 
to a characteristic angular scale $\sim 1^\circ$; this 
corresponds to
the first and strongest peak in the CMB angular power spectrum.
Higher harmonics correspond to modes at other density extrema.
The acoustic oscillations depend on the cosmology as well
as the plasma properties, which set the angular scales of the harmonics
as well as the heights of the features, as well as the correspondence
between the temperature anisotropies and the polarization.
This gives rise to the CMB sensitivity to cosmological parameters.

Precision observations of CMB temperature and polarization fluctuations 
over a wide range of scales now exist, and probe a host of cosmological
parameters. Fortunately for BBN, one of the most robust of these is
the cosmic baryon content, usually quantified in the CMB
literature via the plasma (baryon + electron) density
$\rho_{\rm b}$ written as
the density
parameter $\Omega_{\rm b} \equiv \rho_{\rm b}/\rho_{\rm crit}$ 
where the critical density $\rho_{\rm crit} = 3H_0^2/8\pi G$,
with $H_0 = 100 \ h \ \rm km \, s^{-1} \, Mpc^{-1}$ 
the present Hubble parameter.
This in turn is related to the baryon-to-photon ratio $\eta$ via
\beq
\label{eq:etaconv}
\eta = \frac{\rho_{\rm crit}}{\avg{m} n_\gamma^0} \Omega_B \, ,
\eeq
where
$n_\gamma^0 = 2\zeta(3) T_0^3/\pi^2$ is the present-day photon number
density.
The mean mass per baryon $\avg{m}$ in eq.~(\ref{eq:etaconv})
is roughly the proton mass,
but slightly lower due to the binding energy of helium.
As a result, the detailed conversion depends very mildly on the
\he4 abundance \cite{gary}, 
and we have
\beq 
\eta_{10} = 273.3036\Omega_Bh^2\left( 1 + 7.16958\times 10^{-3}Y_p \right)\left(
  \frac{2.7255 K}{T_\gamma^0} \right)^3 \, .
\label{eqn:obeta}
\eeq

The CMB also encodes the values of
$Y_p$ and $\neff$  at recombination.   
The effect of helium and thus $Y_p$ is to 
set the number of plasma electrons per baryon.
This controls the Thomson scattering mean free path 
$(\sigma_{\rm T} n_e)^{-1}$ that sets the scale at which
the acoustic peaks are damped (exponentially!) by photon diffusion.
The effect of radiation and thus of $\neff$
is predominantly to increase the cosmic expansion rate
via $H^2 \propto \rho$.  This also affects the scale of damping
onset, 
when the diffusion length is comparable to the sound horizon
\cite{hou}.
Thus, measurements of the CMB damping tail at small angular scales
(high $\ell$) probe both $Y_p$ and $\neff$.  In fact, the CMB
determinations of these quantities are strongly anticorrelated--a higher $Y_p$
implies a lower diffusion mean free path, which is equivalent to
a larger sound horizon $\sim \int da/a H$ and thus lower $\neff$.

Since the first {\it WMAP} results in 2003 \cite{wmap1}
sharply measured the first acoustic peaks,
the CMB has determined $\eta$ more precisely than BBN.
Moreover, both ground-based and {\em Planck} data now
measure the damping tail with sufficient accuracy to
simultaneously probe all of $(\Omega_{\rm b} h^2,Y_p,\neff)$
\cite{Planck2015,SPT,ACT}.
Of course, BBN demands an essentially unique relationship among these
quantities.
{\em Thus, it is now possible to meaningfully test BBN and thus cosmology
via CMB measurements alone!}

We adopt CMB  determinations of
$(\Omega_{\rm b} h^2,Y_p,\neff)$ from {\em Planck} 2015 data, 
as described in detail below, \S \ref{sect:cmbpars}.
We note in passing that these CMB constraints rely on 
the present CMB temperature, $T_\gamma^0$, which is
held fixed in the {\em Planck} analysis we use.
In the future, $T_\gamma^0$
should be varied in the CMB data evaluation, in order to provide the
most stringent constraint on the relative number of baryons in the
universe.

\section{The Likelihood Analysis}

As data analysis improved, theoretical studies of BBN moved toward a more rigorous approach
using Monte-Carlo techniques in likelihood analyses \cite{kr,kk,hata,fo}.
Thus, in order to make quantitative statements about the light element
predictions and convolutions with CMB constraints, we need probability
distributions for our BBN predictions, for the light element
observations and CMB-constrained parameters.  We discuss here how we
propagate nuclear reaction rate uncertainties into the BBN light
element abundance predictions, how we determine the CMB-parameter
likelihood distributions, and how we combine them to make stronger
constraints.

\subsection{Monte-Carlo Predictions for the Light Elements}

The dominant source of uncertainty in the BBN light-element
predictions stem from experimental uncertainties of nuclear reaction rates. We
propagate these uncertainties by randomly drawing rates according their
adopted probability distributions for each BBN evaluation.  We choose a Monte
Carlo of size $N=10000$, keeping the error in the mean and error in the error
at the 1\% level.  It is important that we use the same random numbers for
each set of parameters ($\eta, N_\nu$).  This helps remove any extra noise
from the Monte Carlo predictions and allows for smooth interpolations between
parameter points.

For each grid point of parameter values we calculate the means and covariances
of the light element abundance predictions.  We add the $1/\sqrt{N}$ errors in
quadrature to our evaluated uncertainties on the light element predictions. We
have examined the light element abundance distributions, by calculating higher
order statistics (skewness and kurtosis), and by histogramming the resultant
Monte Carlo points and verified that they are well-approximated with
log-normal or gaussian distributions.

\begin{center}
\begin{figure}[htb]
\psfig{file=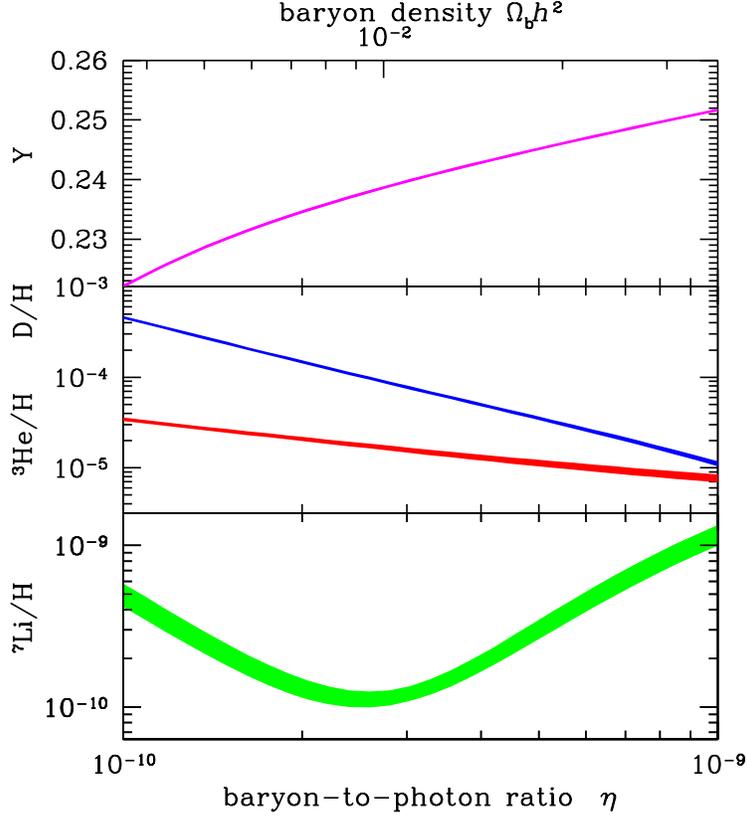,width=0.8\textwidth}
\vskip -1.7in
\caption{
Primordial abundances of the light nuclides as a function of
cosmic baryon content, as predicted by SBBN (``Schramm plot'').
These
results assume $N_\nu = 3$ and the current measurement of the neutron lifetime $\tau_n = 880.3 \pm 1.1$ s. 
Curve widths show $1-\sigma$ errors.
\label{fig:schramm}
}
\end{figure}
\end{center}

In standard BBN, the baryon-to-photon ratio ($\eta$) is the only free
parameter of the theory.  Our Monte Carlo error propagation is
summarized in Figure~\ref{fig:schramm}, which plots the light element abundances
as a function of the baryon density (upper scale) and $\eta$ (lower scale). 
The abundance for He is shown as the mass fraction $Y$, while the abundances of
the remaining isotopes of D, \he3, and \li7 are shown as abundances by number relative to H. 
The thickness of the curves show the $\pm 1 \sigma$ spread in the predicted abundances. These
results assume $N_\nu = 3$ and the current measurement of the neutron lifetime $\tau_n = 880.3 \pm 1.1$ s. 

Using a Monte Carlo approach also allows us to extract sensitivities
of the light element predictions to reaction rates and other
parameters.  The sensitivities are defined as the logarithmic
derivatives of the light element abundances with respect to each
variation about our fiducial model parameters
\cite{fiorentini}, yielding a simple
relation for extrapolating about the fiducial model:
\beq
\label{eqn:sens}
X_i = X_{i,0}\prod_n \left(\frac{p_n}{p_{n,0}}\right)^{\alpha_n} \, ,
\eeq
where $X_i$ represents either the helium mass fraction or the abundances of the other
light elements by number. The $p_n$ represent input quantities to the BBN calculations 
($\eta, N_\nu, \tau_n$) and the gravitational constant $G_N$ as well key nuclear rates which
affect the abundance $X_i$.  $p_{n,0}$ refers to our standard input value. The information contained 
in Eqs. (\ref{yfit}-\ref{li6fit}) are neatly summarized in Table \ref{tab:sens}. 

\beqar
Y_p &=& 0.24703\!\left(\frac{10^{10}\eta}{6.10}\right)^{0.039}\!\!\left(\frac{N_\nu}{3.0}\right)^{0.163}\!\!\left(\frac{G_N}{G_{N,0}}\right)^{0.35}\!\!\left(\frac{\tau_n}{880.3 s}\right)^{0.73} \nonumber \\
 &\times& \left[ p(n,\gamma)d\right]^{0.005}\left[ d(d,n)\he3\right]^{0.006}\left[ d(d,p)t\right]^{0.005} 
\label{yfit}
\eeqar
\beqar
\frac{\rm D}{\rm H} &=& 2.579\!\times\! 10^{-5}\!\left(\frac{10^{10}\eta}{6.10}\right)^{-1.60}\!\!\left(\frac{N_\nu}{3.0}\right)^{0.395}\!\!\left(\frac{G_N}{G_{N,0}}\right)^{0.95}\!\!\left(\frac{\tau_n}{880.3 s}\right)^{0.41} \nonumber  \\
&\times& \left[ p(n,\gamma)d\right]^{-0.19}\left[ d(d,n)\he3\right]^{-0.53}\left[ d(d,p)t\right]^{-0.47} \nonumber \\
&\times& \left[d(p,\gamma)\he3\right]^{-0.31}\left[\he3(n,p)t\right]^{0.023}\left[\he3(d,p)\he4\right]^{-0.012} 
\eeqar
\beqar
\frac{\he3}{\rm H} &=& 9.996\!\times\! 10^{-6}\!\left(\frac{10^{10}\eta}{6.10}\right)^{-0.59}\!\!\left(\frac{N_\nu}{3.0}\right)^{0.14}\!\!\left(\frac{G_N}{G_{N,0}}\right)^{0.34}\!\!\left(\frac{\tau_n}{880.3 s}\right)^{0.15} \nonumber \\
&\times& \left[ p(n,\gamma)d\right]^{0.088}\left[ d(d,n)\he3\right]^{0.21}\left[ d(d,p)t\right]^{-0.27} \nonumber  \\
&\times& \left[d(p,\gamma)\he3\right]^{0.38}\left[\he3(n,p)t\right]^{-0.17}\left[\he3(d,p)\he4\right]^{-0.76}\left[t(d,n)\he4\right]^{-0.009} 
\eeqar
\beqar
\frac{\li7}{\rm H} &=& 4.648\!\times\! 10^{-10}\!\left(\frac{10^{10}\eta}{6.10}\right)^{2.11}\!\!\left(\frac{N_\nu}{3.0}\right)^{-0.284}\!\!\left(\frac{G_N}{G_{N,0}}\right)^{-0.73}\!\!\left(\frac{\tau_n}{880.3 s}\right)^{0.43} \nonumber  \\
&\times& \left[ p(n,\gamma)d\right]^{1.34}\left[ d(d,n)\he3\right]^{0.70}\left[ d(d,p)t\right]^{0.065} \nonumber  \\
&\times& \left[d(p,\gamma)\he3\right]^{0.59}\left[\he3(n,p)t\right]^{-0.27}\left[\he3(d,p)\he4\right]^{-0.75}\left[t(d,n)\he4\right]^{-0.023} \nonumber  \\
&\times& \left[\he3(\alpha,\gamma)\be7\right]^{0.96}\left[\be7(n,p)\li7\right]^{-0.71}\left[\li7(p,\alpha)\he4\right]^{-0.056}\left[t(\alpha,\gamma)\li7\right]^{0.030} 
\eeqar
\beqar
\frac{\li6}{\rm H} &=& 1.288\!\times\! 10^{-13}\!\left(\frac{10^{10}\eta}{6.10}\right)^{-1.51}\!\!\left(\frac{N_\nu}{3.0}\right)^{0.60}\!\!\left(\frac{G_N}{G_{N,0}}\right)^{1.40}\!\!\left(\frac{\tau_n}{880.3 s}\right)^{1.37} \nonumber \\
&\times& \left[ p(n,\gamma)d\right]^{-0.19}\left[ d(d,n)\he3\right]^{-0.52}\left[ d(d,p)t\right]^{-0.46} \nonumber \\
&\times& \left[d(p,\gamma)\he3\right]^{-0.31}\left[\he3(n,p)t\right]^{0.023}\left[\he3(d,p)\he4\right]^{-0.012}\left[d(\alpha,\gamma)\li6\right]^{1.00}
\label{li6fit}
\eeqar

\begin{table}[h]
\caption{This table contains the sensitivities, $\alpha_n$'s defined in Eq.~\ref{eqn:sens} for each of the light element abundance predictions, varied with respect to key parameters and reaction rates.
\label{tab:sens}
}
\begin{tabular}{|l|c|c|c|c|c|}
\hline
Variant &	$Y_p$ &	D/H &	\he3/H & \li7/H & \li6/H \\
\hline
\hline
$\eta$ (6.1$\times 10^{-10}$) &	0.039 &	-1.598 &-0.585 &	2.113 &	-1.512 \\
\hline
$N_\nu$ (3.0) & 0.163 & 0.395 & 0.140 & -0.284 & 0.603 \\
\hline
$G_N$&	0.354&	0.948&	0.335&	-0.727&	1.400 \\
\hline
n-decay&	0.729&	0.409&	0.145&	0.429&	1.372 \\
\hline
p(n,$\gamma$)d&	0.005&	-0.194&	0.088&	1.339&	-0.189 \\
\hline
\he3(n,p)t&	0.000&	0.023	&-0.170&	-0.267&	0.023 \\
\hline
\be7(n,p)\li7&	0.000&	0.000&	0.000&	-0.705&	0.000 \\
\hline
d(p,$\gamma$)\he3&	0.000&	-0.312&	0.375&	0.589&	-0.311 \\
\hline
d(d,$\gamma$)\he4&	0.000&	0.000&	0.000&	0.000&	0.000 \\
\hline
\li7(p,$\alpha$)\he4&	0.000&	0.000&	0.000&	-0.056&	0.000 \\
\hline
d($\alpha,\gamma$)\li6&	0.000&	0.000&	0.000&	0.000&	1.000 \\
\hline
t($\alpha,\gamma$)\li7&	0.000&	0.000&	0.000&	0.030&	0.000 \\
\hline
\he3($\alpha,\gamma$)\be7&	0.000&	0.000&	0.000&	0.963&	0.000 \\
\hline
d(d,n)\he3&	0.006&	-0.529&	0.213&	0.698&	-0.522 \\ 
\hline
d(d,p)t&	0.005&	-0.470&	-0.265&	0.065&	-0.462 \\
\hline
t(d,n)\he4&	0.000&	0.000&	-0.009&	-0.023&	0.000 \\
\hline
\he3(d,p)\he4&	0.000&	-0.012&	-0.762&	-0.752&	-0.012 \\
\hline
\hline
\end{tabular}
\end{table}

\subsection{The Neutron Mean Life}

\begin{figure}[h]
\vskip -2.0in
\psfig{file=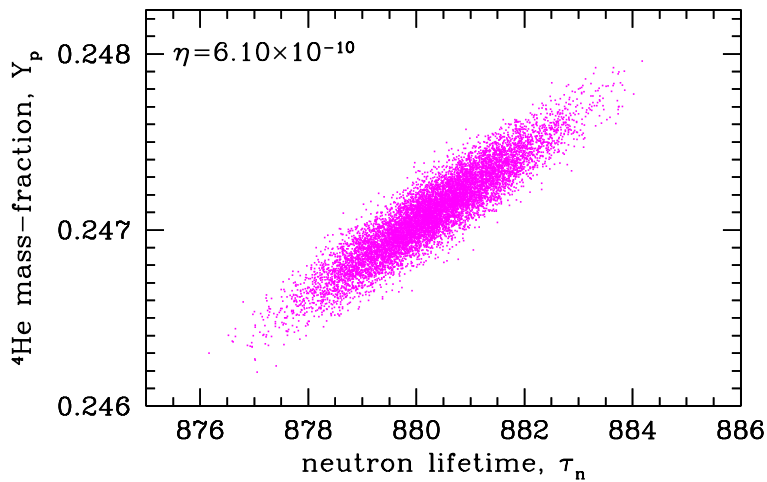,width=0.8\textwidth}
\vskip -1.6in
\caption{
The sensitivity of the \he4 abundance to the neutron mean life, as shown through a scatter plot of our Monte Carlo error propagation.  
\label{fig:Ytau}
}
\end{figure}

As noted in the introduction, the value of the neutron mean life has
had a turbulent history.  Unfortunately, the predictions of SBBN
remain sensitive to this quantity. This sensitivity is displayed in the scatter plot of
our Monte Carlo error propagation with fixed $\eta=6.10\times 10^{-10}$
in Figure~\ref{fig:Ytau}.  The correlation between the neutron mean
lifetime and \he4 abundance prediction is clear.  The correlation is
not infinitesimally narrow because other reaction rate uncertainties
significantly contribute to the total uncertainty in \he4.

\subsection{{\em Planck} Likelihood Functions}
\label{sect:cmbpars}

For this paper, we will need to consider two sets of {\em Planck} Markov Chain
data, one for standard BBN (SBBN) and one for non-standard BBN (NBBN).
Using the {\em Planck} Markov chain data \cite{Planckdata}, we have constructed the
multi-dimensional likelihoods for the following extended parameter
chains, {\bf base\_yhe} and {\bf base\_nnu\_yhe}, for the 
{\tt plikHM\_TTTEEE\_lowTEB} dataset.
As noted earlier, we do not use the {\em Planck} base chain, as it assumes a BBN relationship
between the helium abundance and the baryon density.  

From these 2 parameter sets we have the following 2- and 3-dimensional
likelihoods from the CMB: ${\mathcal L}_{\rm PLA-base\_yhe}(\omega_b,Y_p)$
and ${\mathcal L}_{\rm PLA-base\_nnu\_yhe}(\omega_b,Y_p,N_\nu)$.
The 2-dimensional {\bf base\_yhe} likelihood is well-represented by
a 2D correlated gaussian distribution, with means and standard
deviations for the baryon density and \he4 mass fraction
\beqar
\omega_b & = & 0.022305 \pm0.000225 \\
Y_p & = & 0.25003 \pm0.01367
\eeqar
 and a correlation coefficient 
$r \equiv
{\rm cov}(\omega_b,Y_p)/\sqrt{{\rm var}(\omega_b){\rm var}(Y_p)}=+0.7200$.  

The two parameter data can be marginalized to yield 1-dimensional likelihood functions for
$\eta$. The peak and 1-$\sigma$ spread in $\eta$ is given in the first row of Table \ref{tab:eta}.
The following rows correspond to different determinations of $\eta$. In the second-fourth rows,
no CMB data is used. That is, we fix $\eta$ only from the observed abundances of \he4, D or both.
Notice for example, in row 2, the value for $\eta$ is low and has a huge uncertainty.  This is due to the 
slightly low value for the observational abundance (\ref{ypr}) and the logarithmic dependence of
$Y_p$ on $\eta$. We see again that BBN+$Y_p$ is a poor baryometer. 
This will be described in more detail in the following subsection.  Row 5, uses the BBN relation 
between $\eta$ and $Y_p$, but no observational input from $Y_p$ is used. This is closest to the
{\em Planck} determination found in \cite{Planck2015}, though here $Y_p$ was taken to be free
and the value of $\eta$ in the Table is a result of marginalization over $Y_p$. 
This accounts for the very small difference in the results for $\eta$: $\eta_{10} = 6.09$ ({\em Planck}); 
$\eta_{10} = 6.10$ (Table \ref{tab:eta}). Rows 6-8 add the observational determinations of 
\he4, D and the combination. As one can see, the inclusion of the observational data
does very little to affect the determination of $\eta$ and thus we use $\eta_{10} = 6.10$ 
as our fiducial baryon-to-photon ratio.

\begin{table}[ht]
\caption{Constraints on the baryon-to-photon ratio, using different combinations of observational constraints.  We have marginalized over $Y_p$ to create 1D $\eta$ likelihood distributions.
\label{tab:eta}
}
\begin{tabular}{|l|c|}
\hline
 Constraints Used & $\eta\times 10^{10}$ \\
\hline
CMB-only & $6.108\pm 0.060$ \\
\hline
\hline
BBN+$Y_p$ & $4.87^{+2.46}_{-1.54}$ \\
\hline
BBN+D & $6.180\pm 0.195$ \\
\hline
BBN+$Y_p$+D & $6.172\pm 0.195$  \\
\hline
CMB+BBN & $6.098\pm 0.042$ \\
\hline
CMB+BBN+$Y_p$ & $6.098\pm 0.042$ \\
\hline
CMB+BBN+D & $6.102\pm 0.041$ \\
\hline
\hline
CMB+BBN+$Y_p$+D & $6.101\pm 0.041$ \\
\hline
\end{tabular}
\end{table}

The 3-dimensional {\bf base\_nnu\_yhe} likelihood is {\em not}
well-represented by a simple 3D correlated gaussian distribution, but
since these distributions are single-peaked we can correct for the
non-gaussianity via a 3D Hermite expansion about a 3D correlated
gaussian base distribution. Details of this prescription will be given in the Appendix.

The calculated mean values and standard
deviations for these distributions are: 
\beqar
\omega_b &=& 0.022212\pm0.000242 \\
\neff &=& 2.7542\pm0.3064  \label{3dneff}\\
Y_p &=& 0.26116\pm0.01812
\eeqar
These values correspond to the peak of the likelihood distribution using CMB data alone. 
That is, no use is made of the correlation between the baryon density and the helium abundance through BBN.
For this reason, the helium mass fraction is found to be rather high. Our value of $Y_p = 0.261 \pm 0.36 (2\sigma)$ can be compared with the value given by the {\em Planck} collaboration \cite{Planck2015} of 
$Y_p = 0.263^{+0.34}_{-0.37}$. 

In this case, we marginalize to form a 2-d likelihood function to determine both $\eta$ and $\neff$. 
As in the 1-d case discussed above, we can determine $\eta$ and $N_\nu$ using CMB data
alone.  This result is shown in row 1 of Table \ref{tab:etannu} and does not use
any correlation between $\eta$ and $Y_p$. Note that the value of $N_\nu$ given here
differs from that in Eq. (\ref{3dneff}) since the value in the Table comes from a marginalized likelihood function, where as the value in the equation does not.  Row 2, uses only BBN and the observed abundances
of \he4 and D with no direct information from the CMB.  Rows 3-6 use the combination of the
CMB data, together with the BBN relation between $\eta$ and $Y_p$ with and without the observational
abundances as denoted. As one can see, opening up the parameter space to allow $N_\nu$ to float
induces a relatively small drop $\eta$ (by a fraction of 1 $\sigma$) and the peak for $N_\nu$ is below
the Standard Model value of 3 though consistent with that value within 1 $\sigma$. 

\begin{table}[ht]
\caption{The marginalized most-likely values and central 68.3\% confidence limits on the baryon-to-photon ratio and effective number of neutrinos, using different combinations of observational constraints.
\label{tab:etannu}
}
\begin{tabular}{|l|c|c|}
\hline
 Constraints Used & $\eta_{10}$ & $N_\nu$ \\
\hline
CMB-only & $6.08\pm0.07$ & $2.67^{+0.30}_{-0.27}$ \\
\hline
BBN+$Y_p$+D & $6.10\pm0.23$  & $2.85\pm0.28$ \\
\hline
CMB+BBN & $6.08\pm0.07$ & $2.91\pm0.20$ \\
\hline
CMB+BBN+$Y_p$ & $6.07\pm 0.06$ & $2.89\pm 0.16$ \\
\hline
CMB+BBN+D & $6.07\pm0.07$ & $2.90\pm0.19$ \\
\hline
CMB+BBN+$Y_p$+D & $6.07\pm 0.06$ & $2.88\pm0.16$ \\
\hline
\end{tabular}
\end{table}

We note that we have been careful to use the appropriate relation between $\eta$ and
$\omega_b$ via Eq.~\ref{eqn:obeta}.  Also, in our NBBN calculations we
formally use the number of neutrinos, {\em not} the {\em effective} number
of neutrinos, thus demanding the relation:
$\neff=1.015333N_\nu$. For the 2D {\bf base\_yhe} CMB likelihoods,
we include the higher order skewness and kurtosis terms to
more accurately reproduce the tails of the distributions.

\subsection{Results:  The Likelihood Functions}

Applying the formalism described above, we derive the likelihood functions
for SBBN and NBBN that are our central results.
Turning first to SBBN, we fix $N_\nu = 3$
and use the {\em Planck} determination of $\eta$ as
the sole {\em input} to BBN in order to derive CMB+BBN 
predictions for each light element. 
That is, for each light element species $X_i$ we
evaluate the likelihood
\beq
{\mathcal L}(X_i) \propto \int 
  {\mathcal L}_{\rm PLA-base\_yhe}(\omega_b,Y_p) \
  {\mathcal L}_{\rm BBN}(\eta;\{X_i\}) \ d\eta
\eeq
where 
${\mathcal L}_{\rm BBN}(\eta;\{X_i\})$ comes from our BBN
Monte Carlo, and where we use the $\eta-\omega_{\rm b}$ relation
in eq.~(\ref{eqn:obeta}).
In the case of \he4, we use only the CMB $\eta$ to determine
the $X_i = Y_{p,\rm BBN}$ prediction and compare this to the CMB-only prediction.

\begin{center}
\begin{figure}[htb]
\psfig{file=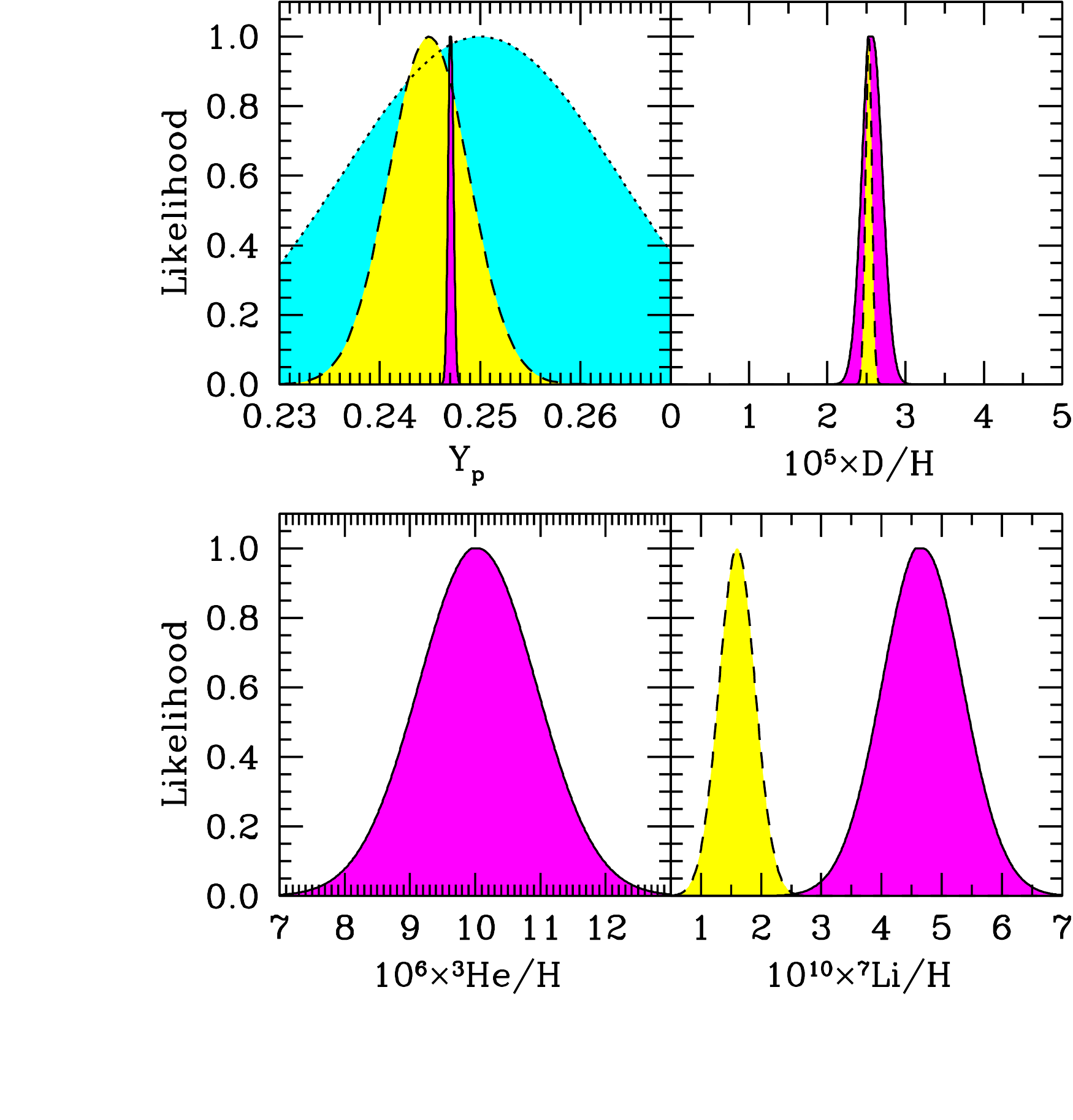,width=0.75\textwidth}
\caption{
Light element predictions using the CMB determination of the 
cosmic baryon density.  Shown are likelihoods for each of the light nuclides,
normalized to show a maximum value of 1.  The solid-lined,
dark-shaded (purple) curves are 
the BBN+CMB predictions, based on {\em Planck} inputs as discussed
in the text.  The dashed-lined, light-shaded 
(yellow) curves show astronomical measurements
of the primordial abundances, for all but \he3 where 
reliable primordial abundance measures do not exist.  For 
\he4, the dotted-lined, medium-shaded (cyan) curve shows
the CMB determination of \he4.
\label{fig:LYi}
}
\end{figure}
\end{center}

The resulting CMB+BBN abundance likelihoods appear 
as the dark-shaded (purple, solid line) curves in Figure \ref{fig:LYi},
which also shows the observational abundance constraints
(\S \ref{sect:obs}) in the light-shaded (yellow, dashed-line)
curves.
In panel (a), we see that the \he4 BBN+CMB likelihood is
markedly more narrow than its observational counterpart,
but the two are in near-perfect agreement.
The medium-shaded (cyan, dotted line) curve in this panel
is the CMB-only $Y_p$ prediction,
which is the least precise but also completely consistent
with the other distributions.
Panel (b) displays the dramatic consistency between
the CMB+BBN deuterium prediction and the observed high-$z$ abundance.
Moreover, we see that the D/H observations 
are substantially more precise than the theory.
Panel (c) shows the primordial \he3 prediction, for which there is 
no reliable observational test at present.
Finally, panel (d) reveals a sharp discord between
the BBN+CMB prediction for \li7 and the observed primordial
abundance--the two likelihoods are essentially disjoint.

Figure \ref{fig:LYi} represents not only a quantitative assessment of the 
concordance of BBN, but also a test of the
standard big bang cosmology.
If we limit our attention to each element in turn,
we are struck by the spectacular agreement between
D/H observations at $z \sim 3$ and the BBN+CMB predictions
combining physics at $z \sim 10^{10}$ and $z \sim 1000$.
The consistency among all three $Y_p$ determinations is 
similarly remarkable, and the joint concordance
between D and \he4 represents a non-trivial success of the
hot big bang model.  Yet this concordance is not complete:
the pronounced discrepancy in \li7 measures 
represents the ``lithium problem'' discussed below
(\S \ref{sect:Liprob}).  This casts a shadow of doubt
over SBBN itself, pending a firm resolution of the lithium problem,
and until then the BBN/CMB concordance remains an
incomplete success for cosmology.

Quantitatively, the likelihoods in Fig.~\ref{fig:LYi}
are summarized by
the predicted abundances 
\beqar
Y_p &=& 0.24709\pm0.00025 \\
{\rm D/H} &=& (2.58\pm0.13)\times 10^{-5} \\
\he{3}/{\rm H} &=& (10.039\pm0.090)\times 10^{-5} \\
\li{7}/{\rm H} &=& (4.68\pm0.67)\times 10^{-10} \\
\log_{10}{\left( \li{6}/{\rm H} \right)} &=& -13.89\pm0.20
\eeqar
where the central value give the mean,
and the error the $1\sigma$ variance.
The slightly differences from the values in
Table \ref{tab:re} arise due to the Monte Carlo averaging procedure here as opposed
to evaluating the abundance using central values of all inputs
at a single $\eta$.

\begin{center}
\begin{figure}[htb]
\vskip -.5in
\psfig{file=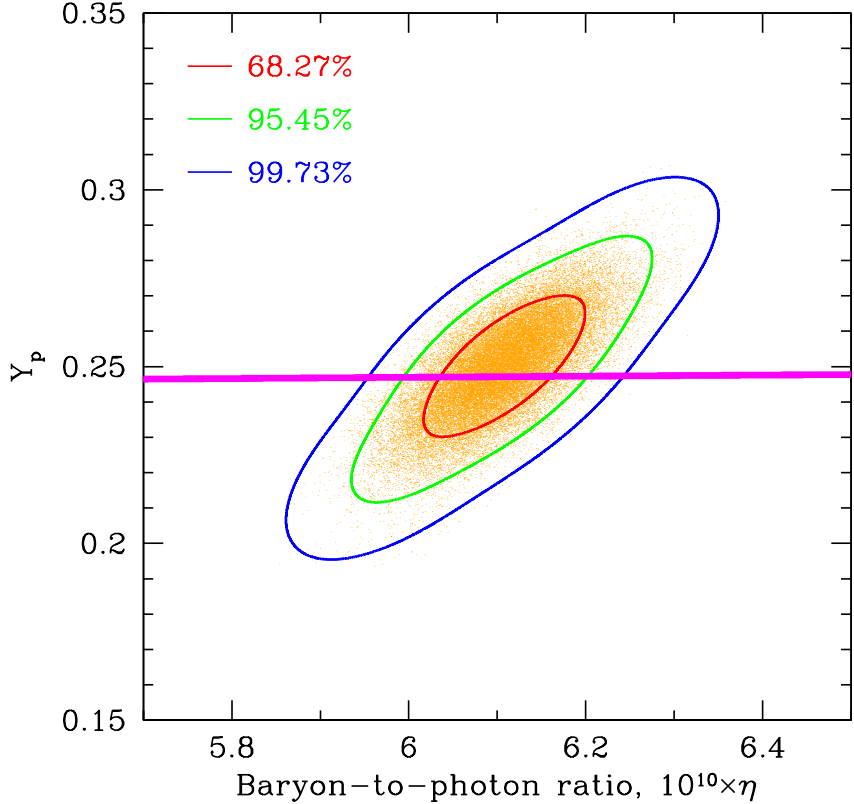,width=0.8\textwidth}
\vskip -1.4in
\caption{The 2D likelihood function contours derived from the {\em Planck} 
Markov Chain Monte Carlo {\bf base\_yhe}  \cite{Planckdata}
with fixed $N_\nu = 3$ (points).  
The correlation between $Y_p$ and $\eta$ is evident.
The 3-$\sigma$ BBN prediction for the helium mass fraction is shown with the colored band.
We see that including the BBN $Y_p(\eta)$ relation significantly reduces
the uncertainty in $\eta$ due to the CMB $Y_p-\eta$ correlation.
\label{fig:L2D1}
}
\end{figure}
\end{center}

We see that the BBN/CMB comparison is enriched now that
the CMB has achieved an interesting sensitivity to $Y_p$
as well as $\eta$.
This interplay is further illustrated in 
Figure \ref{fig:L2D1}, which shows 2-D likelihood contours
in the $(\eta,Y_p)$ plane, still for fixed $N_\nu = 3$.
The {\em Planck} contours show a positive correlation between
the CMB-determined baryon density and helium abundance.
Also plotted is the BBN relation for $Y_p(\eta)$, which
for SBBN is a {\em zero-parameter curve}
that is very tight even 
including its small width due to nuclear reaction rate
uncertainties.  
We see that the curve
goes through the heart of the CMB predictions,
which represents a novel and non-trivial test of SBBN
{\em based entirely on CMB data} without any 
astrophysical input.
This agreement stands as a triumph for SBBN and the hot big bang,
and illustrates the still-growing power of the CMB
as a cosmological probe.

\begin{center}
\begin{figure}[htb]
\vskip -.7in
\psfig{file=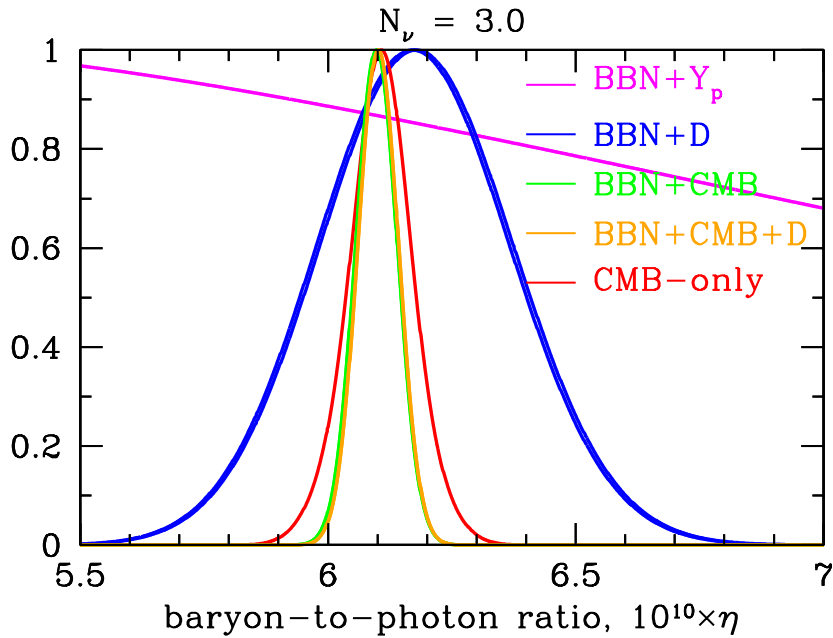,width=1\textwidth}
\vskip -1.8in
\caption{
The likelihood distributions of the baryon-to-photon ratio parameter, $\eta$, given various CMB and light-element abundance constraints.
\label{fig:LEta1}
}
\end{figure}
\end{center}

Thus far we have used the CMB $\eta$ as an input to BBN;
we conclude this section by studying the constraints on $\eta$
when jointly using BBN theory, light-element abundances, and the CMB
in various combinations. 
Figure \ref{fig:LEta1} shows the $\eta$ likelihoods
that result from a set of such combinations.
Setting aside at first the CMB,
the BBN+$X$ curves show the combination of BBN theory
and astrophysical abundance observations,
${\mathcal L}_{{\rm BBN}+X}(\eta) = \int {\mathcal L}_{\rm BBN}(\eta,X) \ 
{\mathcal L}_{\rm obs}(X) \ dX$, with $X \in (Y_p,{\rm D/H})$.
The CMB-only curve marginalizes over the {\em Planck} $Y_p$ values
${\mathcal L}_{\rm CMB-only}(\eta) = 
\int {\mathcal L}_{\rm PLA-base\_yhe}(\omega_b,Y_p) \ dY_p$
where we use the $\eta-\omega_{\rm b}$ relation
in eq.~(\ref{eqn:obeta}).
The BBN+CMB curve adds the BBN $Y_p(\eta)$ relation.
Finally, BBN+CMB+D also includes the observed primordial deuterium.

We see in Fig.~\ref{fig:LEta1} that of the primordial abundance observations,
deuterium is the only useful ``baryometer,'' due to its
strong dependence on $\eta$ in the Schramm plot (Fig.~\ref{fig:schramm}).
By contrast, \he4 alone offers no useful
constraint on $\eta$, tracing back to the weak $Y_p(\eta)$ trend
in Fig.~\ref{fig:schramm}.
The CMB alone has now surpassed BBN+D in measuring 
the cosmic baryon content, but an even stronger
limit comes from BBN+CMB. 
As seen in Fig.~\ref{fig:L2D1}, 
this tightens the $\eta$ constraint due to
the CMB correlation between $Y_p$ and $\eta$
Finally BBN+CMB+D provides only negligibly stronger
limits.
The peaks of the likelihoods correspond to the values
in Table \ref{tab:eta}, and the tightest constraints are all consistent
with our adopted central value $\eta = 6.10 \times 10^{-10}$.

\section{The Lithium Problem}
\label{sect:Liprob}

As seen in the panels of Fig.~\ref{fig:LYi} above, the observed
primordial lithium abundance 
differs sharply from the BBN+CMB prediction \cite{cfo5}.
This discrepancy constitutes the
``Lithium Problem'', which was foreshadowed before CMB determinations
of $\eta$, and has persisted over the dozen years 
since the first {\em WMAP} data release.
For a detailed recent review of the lithium problem, see \cite{fieldsliprob}. Here
we briefly summarize the current status.

The most conventional means to resolve the
primordial lithium problem invokes large lithium depletion in halo stars \cite{dep}.
As noted above (\S \ref{sect:LiObs}), recent observations
of the Spite plateau ``meltdown'' at very low metallicity, $[{\rm Fe/H}] < -3$,
seem to demand that {\em some} stars have depleted their lithium \cite{aoki2009,sbordone2010}.
Could the other plateau halo stars have also
destroyed their lithium?
Such a scenario cannot be ruled out, but raises other questions
that remain unanswered:
why is the Li/H dispersion so small at metallicities
above the ``meltdown''?  And why is there a 
``lithium desert'' with no stars having lithium abundances
between the plateau and the primordial abundance?

It is worthwhile to find other
sites for Li/H measurements, as clearly
halo star lithium depletion is theoretically complex
and observationally challenging. 
Unfortunately, the CMB itself does not yet provide an observable
signature of primordial lithium \cite{Switzer}.
However, a promising new direction is the observation of
interstellar lithium in low-metallicity or
high-$z$ galaxies \cite{fkf}.
Interstellar measurements in the Small Magellanic Cloud (metallicity $\sim 1/4$ solar)
find ${\rm Li/H}_{\rm ISM,SMC} = (4.8 \pm 1.8)  \times 10^{-10}$
\cite{howk}.
This value is consistent with the CMB+BBN primordial abundance,
but the SMC is far from primordial, with a metallicity of about 1/4 solar.
Indeed, the SMC interstellar lithium abundance {\em agrees} with that of
Milky Way stars at the same $[{\rm Fe/H}]$, which are disk (Population I)
stars in which Li/H is rising from the Spite plateau due to Galactic 
production.  Thus we see consistency between lithium abundances
at the same metallicity, but
measured in very different systems with very different systematics.
This strongly suggests that stellar lithium depletion has not been
underestimated, at least down to this metallicity.
Moreover, this observation serves as a proof-of-concept demonstration
that measurements of interstellar lithium in galaxies with lower metallicities
would could strongly test stellar depletion and potentially rule out
this solution to the lithium problem.

Another means of resolving the lithium problem 
within the context of the standard cosmology and
Standard Model 
microphysics is to alter the BBN theory predictions due to
revisions in nuclear reaction rates \cite{coc3,cfo4,boyd}.
But as we have seen, all of the reactions that are ordinarily the
most important for BBN have been well measured at the energies of interest.
Typically, cross sections are known to $\sim 10\%$ or better,
and these errors are already folded into Fig.~\ref{fig:LYi}.
A remaining possibility is that a reaction thought to be 
unimportant could contain a {\em resonance}
heretofore unknown, which could boost its cross section enormously,
analogously to the celebrated Hoyle \iso{C}{12} resonance
that dominates the $3\alpha \rightarrow \iso{C}{12}$ rate
\cite{Hoyle}.

In BBN, the densities and timescales prior to nuclear freezeout
are such that only two-body reactions are important,
and it is possible to systematically study all two-body
reactions that enhance the destruction of \be7.
A small number candidates emerge, for which one can make definite
predictions of the needed resonant state energy and width:
$\be7(d,\gamma)\iso{B}{9}$,
$\be7(\he3,\gamma)\iso{C}{10}$,
and $\be7(t,\gamma)\iso{B}{10}$
\cite{cp,chfo,brog}. 
However, measurements in $\be7(d,d)\be7$ \cite{OMalley},
$\iso{Be}{9}(\he3,t)\iso{B}{9}$ 
\cite{Kirsebom},
and an $R$-matrix analysis of \iso{B}{9} \cite{paris}
all rule out a $\iso{B}{9}$ resonance.
Similarly,
 \iso{C}{10} data rule out the needed resonance in $\iso{C}{10}$ \cite{Hammache}.
The upshot is that a ``nuclear option'' to the lithium
problem is essentially excluded.

It is thus a real possibility that the lithium problem may point
to new physics at play during or after nucleosynthesis.
A number of possible solutions have been proposed and are discussed
in the reviews cited above.  Here we simply note that a challenge
to all such models is that they must reduce \li7 
substantially, yet {\em not} perturb the other light elements unacceptably.
Generally, there is a tradeoff between
\be7 destruction and D production (usually as a by-product of \he4 disruption) \cite{dli,opvs}.
Essentially all successful models drive D/H to the maximum abundance
allowed by observations.  However, the new very precise D/H measurements
(\S \ref{sect:deut}) dramatically reduce the allowed perturbations
and will challenge most of the existing new-physics solutions to the 
lithium problem.  It remains to be seen whether it is possible to introduce
new physically-motivated perturbations that satisfy the D/H 
constraint while still solving or at least substantially reducing the lithium
problem.

\section{Limits on $\neff$}
\label{sect:Neff}

\begin{center}
\begin{figure}[htb]
\psfig{file=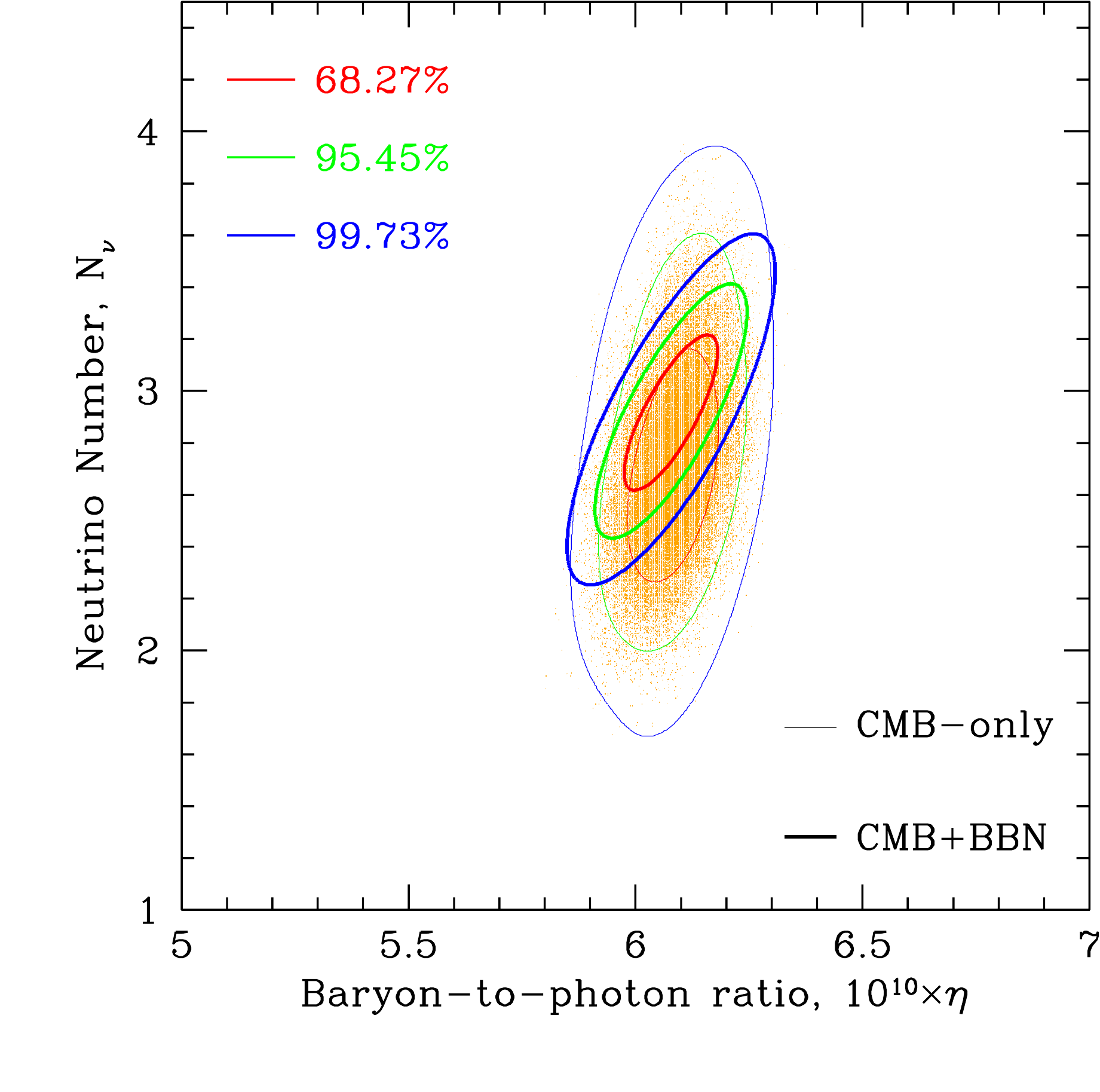,width=0.75\textwidth}
\caption{The 2D likelihood function contours derived from the
{\em Planck} Markov Chain Monte Carlo {\bf base\_nnu\_yhe} \cite{Planckdata}, marginalized
over the CMB $Y_p$ (points).
Thin contours are for CMB data only, while thick contours use the
BBN $Y_p(\eta)$ relation,
assuming no observational constraints on the light elements.
We see that that whereas in the CMB-only case $N_\nu$ and $\eta$ 
are almost uncorrelated, 
in the CMB+BBN case a stronger correlation emerges.
\label{fig:L2Detanu}
}
\end{figure}
\end{center}

Before concluding, we consider a one-parameter extension of SBBN
by allowing the number of relativistic degrees of freedom to differ from 
the Standard Model value of $N_\nu = 3$ and $\neff = 3.046$. 
Opening this degree of freedom has an impact on both the CMB and BBN.
In Fig.~\ref{fig:L2Detanu}, the thinner contours 
show the 2D likelihood distribution in the $(\eta,N_\nu)$ plane,
using {\em Planck} data  marginalizing over the CMB $Y_p$.
We see that the CMB $N_\nu$ values are nearly uncorrelated with $\eta$.
The thicker contours include BBN information and are discussed below.

\begin{center}
\begin{figure}[htb]
\vskip -.6in
\psfig{file=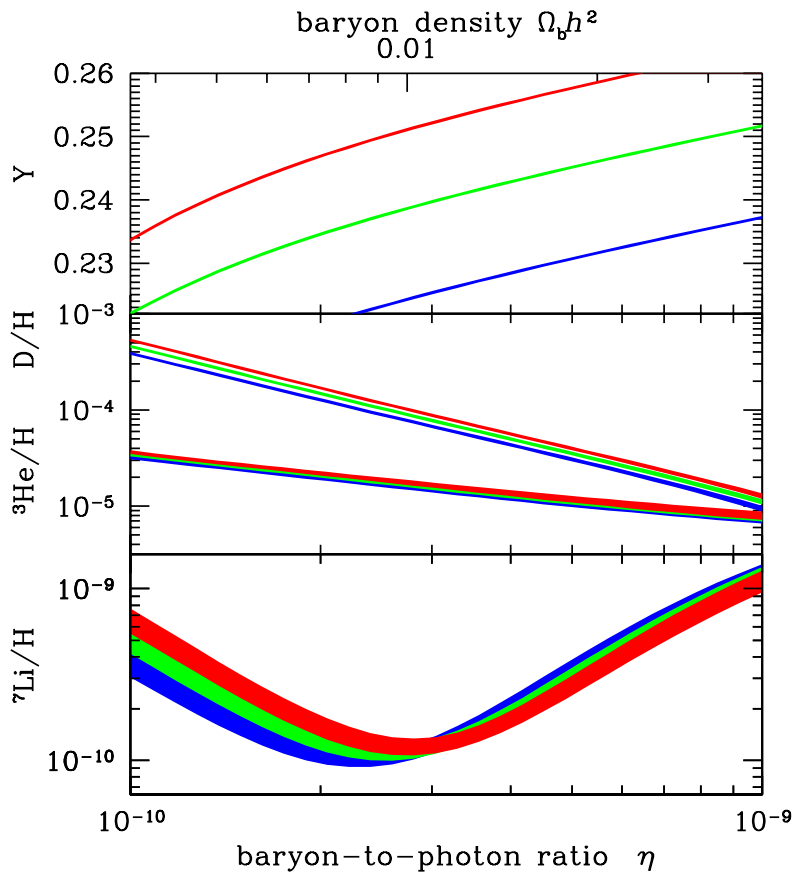,width=0.85\textwidth}
\vskip -1.5in
\caption{
The sensitivity of the light element predictions to the number of neutrino species, similar to Figure~\ref{fig:schramm}.  Here, abundances shown by blue, green, and red bands correspond to 
calculated abundances assuming $N_\nu = 2, 3$ and 4 respectively.
\label{fig:SchrammNnu}
}
\end{figure}
\end{center}

Turning to the effects of $N_\nu$ on BBN, 
eqs. (\ref{rho}) -- (\ref{hubb}) show that
increasing the number of neutrino flavors leads to an increased Hubble parameter which in turn leads to 
an increased freeze-out temperature, $T_f$. Since the neutron-to-proton ratio at freeze-out scales 
as $(n/p) \simeq e^{-\Delta m/T_f}$, higher $T_f$ leads to higher $(n/p)$ and thus higher $Y_p$ \cite{nnu}.
As a consequence, one can establish an upper bound to the number of neutrinos \cite{ssg} if 
in addition one has a lower bound on the baryon-to-photon ratio \cite{ossty} as the helium abundance
also scales monotonically with $\eta$.  The dependence of the helium mass fraction $Y$ on 
both $\eta$ and $N_\nu$ can be seen in Figure \ref{fig:SchrammNnu}
where we see the calculated value of $Y$ for $N_\nu = 2, 3$ and 4 as depicted by the blue, green 
and red curves respectively. In the Figure, one clearly sees not only the monotonic growth of $Y$
with $\eta$, but also the strong sensitivity of $Y$ with $N_\nu$. The importance of a lower 
bound on $\eta$ (or better yet fixing $\eta$) is clearly apparent in setting an upper bound on $N_\nu$. 
Prior to CMB determinations of $\eta$, the lower bound on $\eta$ could be set using a combination
of D and \he3 observations enabling a limit of $N_\nu < 4$ \cite{ytsso} given the estimated uncertainties in $Y_p$ at the time. More aggressive estimates of an upper bound on the helium mass fraction led to tighter
bounds on $N_\nu$ \cite{bbn,tight,kk}. The bounds on $N_\nu$ became more rigorous
when likelihood techniques were introduced \cite{kk,ot,lisi,bbnt,cfo2,cfos,ms}.

While the dependence of $Y_p$ on $N_\nu$ is well documented, 
we also see from Fig. \ref{fig:SchrammNnu}, there is a non-negligible effect on D and \li7
from changes in $N_\nu$ \cite{cfo2}. In particular, while the sensitivity of D to $N_\nu$ is not 
as great as that of $Y_p$, the deuterium abundance is measured much more accurately
and as a result the constraint on $N_\nu$ is now due to both abundance determinations 
as can be discerned from Table \ref{tab:etannu}.

\begin{figure}[htb]
\vskip  -1.8in
\hskip -.75in
\psfig{file=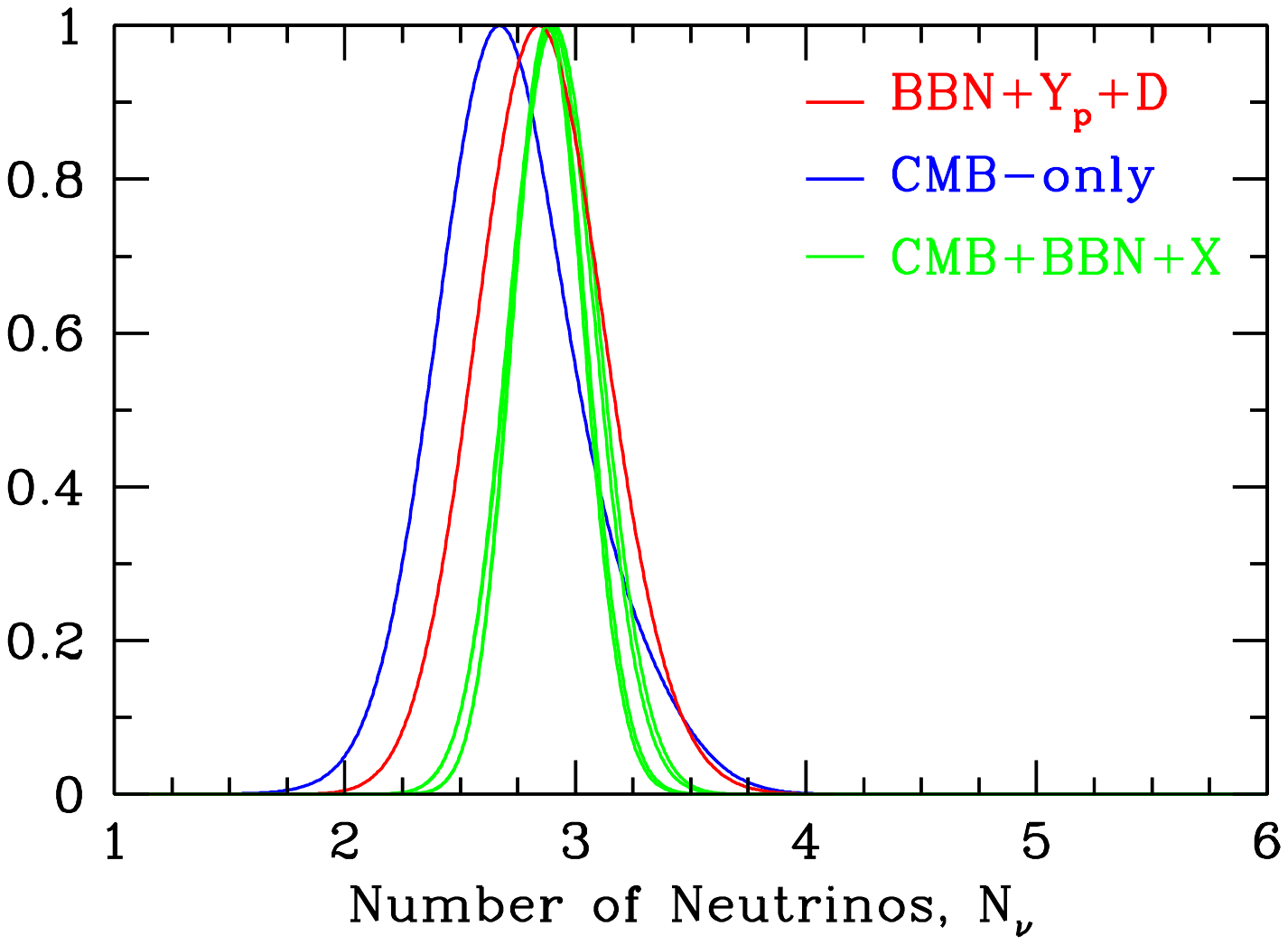,width=0.65\textwidth}
\hskip  -1.25in
\psfig{file=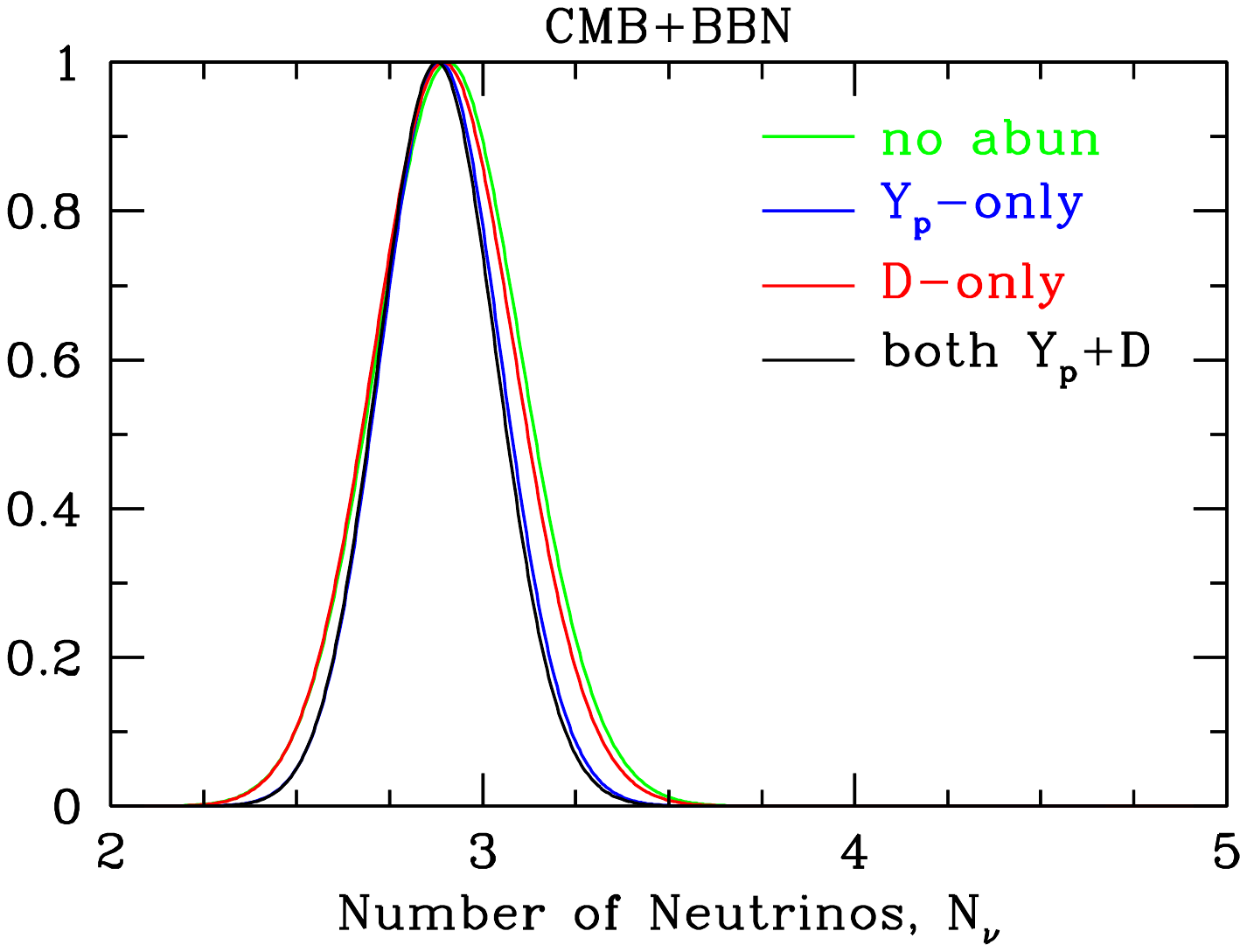,width=0.65\textwidth}
\vskip -1.4in
\caption{
The marginalized distributions for the number of neutrinos, given different combinations of observational constraints. The left panel shows the likelihood function the case where only CMB data is used (blue),
only BBN and abundance data is used (red) and when a combination of BBN and CMB data is used
(green). The four green curves are shown again in the right panel for better clarity.
\label{fig:LNnu}
}
\end{figure}

By marginalizing over the baryon density, we can form 1-d likelihood functions for $N_\nu$.
These are shown in Figure \ref{fig:LNnu}. In the left panel, we show the CMB-only result
by the blue curve. Recall that this uses no BBN correlation between the baryon density and helium
abundance.  While the peak of the likelihood for this case is lowest of the cases considered
($N_\nu = 2.67$) its uncertainty ($\approx 0.30$) makes it consistent with the Standard Model. 
The position of the peak of the likelihood function is given in Table \ref{tab:etannu} for
this case as well as the other cases considered in the Figures. 
In contrast the red curve shows the limit we obtain purely from matching the BBN calculations
with the observed abundances of helium and deuterium.  In this case, the fact that the peak of 
the likelihood function is at $N_\nu = 2.85$ can be traced directly to the fact that the 
central helium abundance is $Y_p = 0.2449$.  Given the sensitivity of $Y_p$ to $N_\nu$
found in Eq. \ref{yfit}, the drop in $N_\nu$ from the Standard Model value of 3.0,
compensates for a helium abundance below the Standard Model prediction closer to 0.247.
Nevertheless, the uncertainty again places the Standard Model within 1 $\sigma$ of the distribution peak. 
The remaining cases displayed (in green) correspond to combining the CMB data with BBN.
There are 4 green curves in the left panel and these have been isolated in the right panel for
better clarity.  As one can see, once one combines the BBN relation between helium and the baryon
density, the actual abundance determinations have only a secondary effect in determining $N_\nu$ which
takes values between 2.88 and 2.91. Using the CMB, BBN and the abundances of both D and \he4,
yields the tightest constraint on the number of neutrino flavors $N_\nu = 2.88 \pm 0.16$, again 
consistent with the Standard Model\footnote{Of all the cases considered, the one that can best be compared with the results presented by the {\em Planck} collaboration \cite{Planck2015} is the case CMB+BBN+D. We find
$\neff = 2.94 \pm 0.38 (2 \sigma)$ while they quote $\neff = 2.91 \pm 0.37$. While we obtain similar results to other cases, direct comparison is complicated not only by the slight difference in the $\eta-Y$ relation due
to different BBN codes, but also by the adopted value for primordial \he4.}. It is interesting to note, that because of the drop in $Y_p$
in the most recent analysis \cite{aos4}, the 95\% CL upper limit on $N_\nu$ is 3.20.

\begin{figure}[htb!]
\vskip  -1.8in
\hskip -.75in
\psfig{file=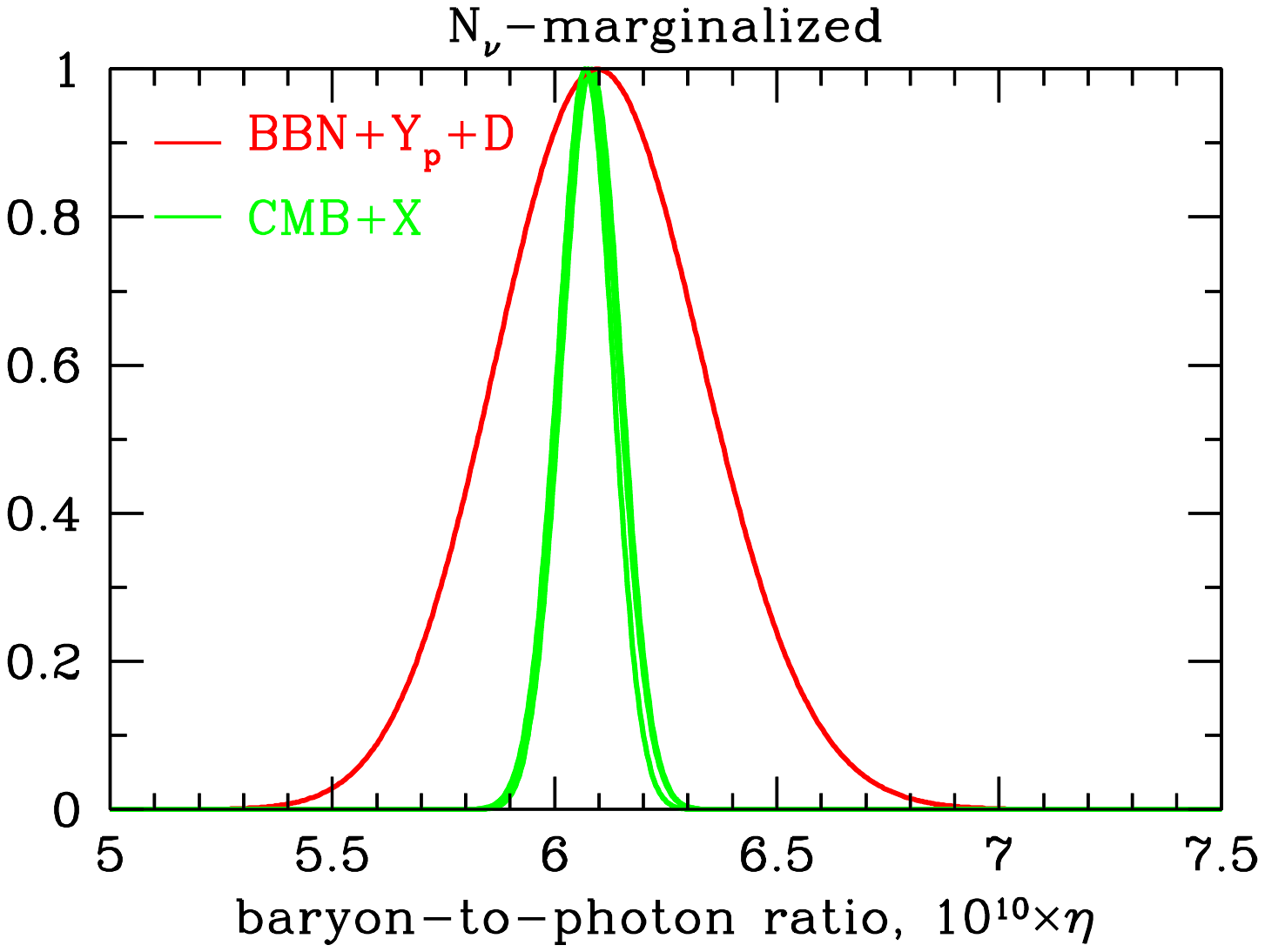,width=0.65\textwidth}
\hskip  -1.25in
\psfig{file=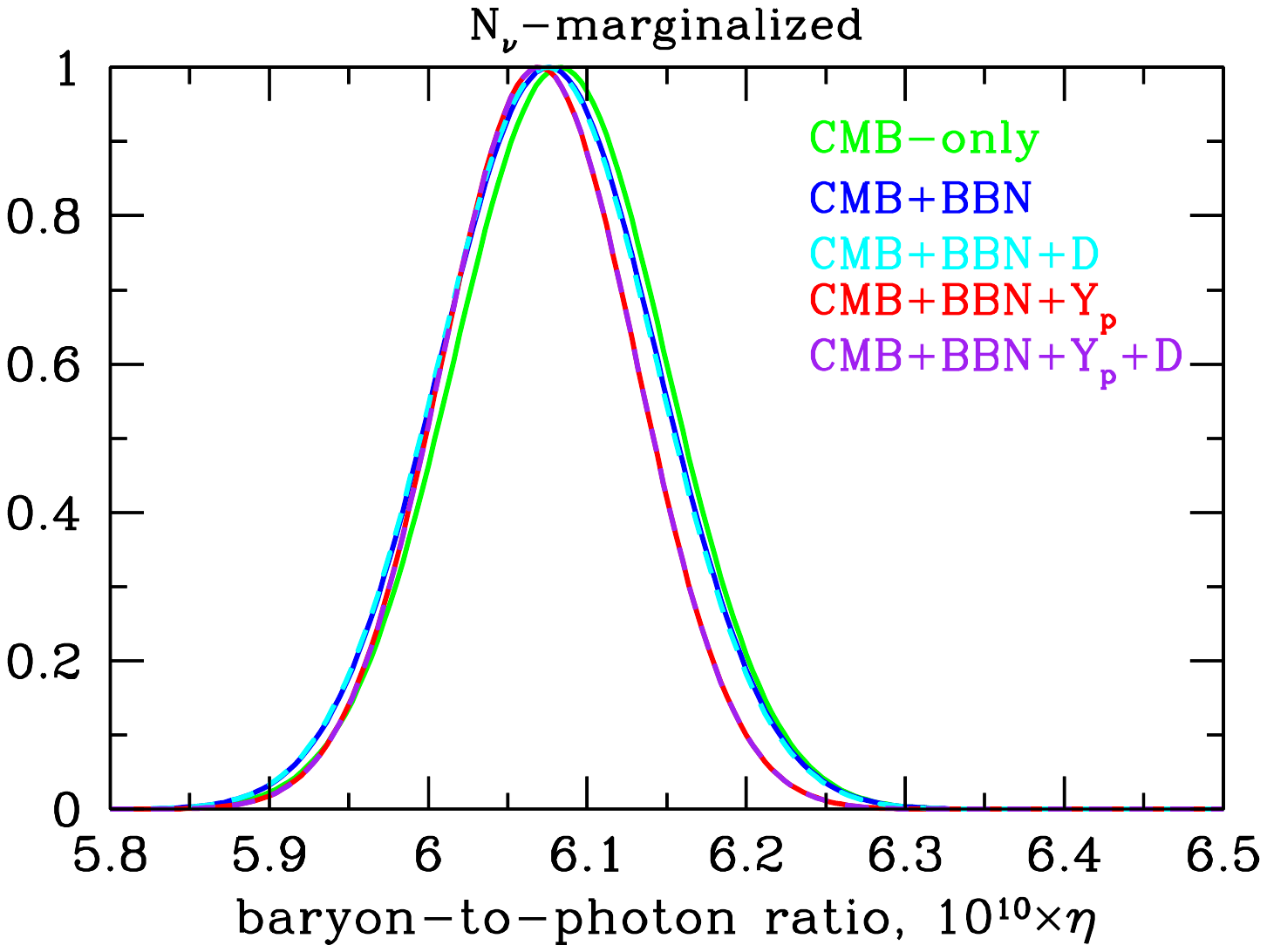,width=0.65\textwidth}
\vskip -1.4in
\caption{
The marginalized distributions for the baryon to photon ratio ($\eta$), given different combinations of observational constraints.
\label{fig:LEta2}
}
\end{figure}

It is also possible to marginalize over the number of neutrino flavors and produce a 1-d likelihood 
function for $\eta_{10}$ as shown in Figure \ref{fig:LEta2}. In the left panel, the
broad distribution shown in red corresponds to the BBN plus abundance data constraint
using no information from the CMB.  Here the baryon density is primarily determined by the D/H abundance.
When the CMB is added, the uncertainty in $\eta$ drops dramatically (from 0.23 to 0.06 or 0.07)
independent of whether abundance data is used. The 5 green curves are almost indistinguishable
and are shown in more detail in the right panel. Once again the peak of the likelihood distributions are given
in Table \ref{tab:etannu}. The values of $\eta$ are slightly lower than the Standard Model results discussed
above. This is due to the additional freedom in the likelihood distribution afforded by the additional parameter,
$N_\nu$.

\begin{center}
\begin{figure}[htb!]
\vskip -.5in
\psfig{file=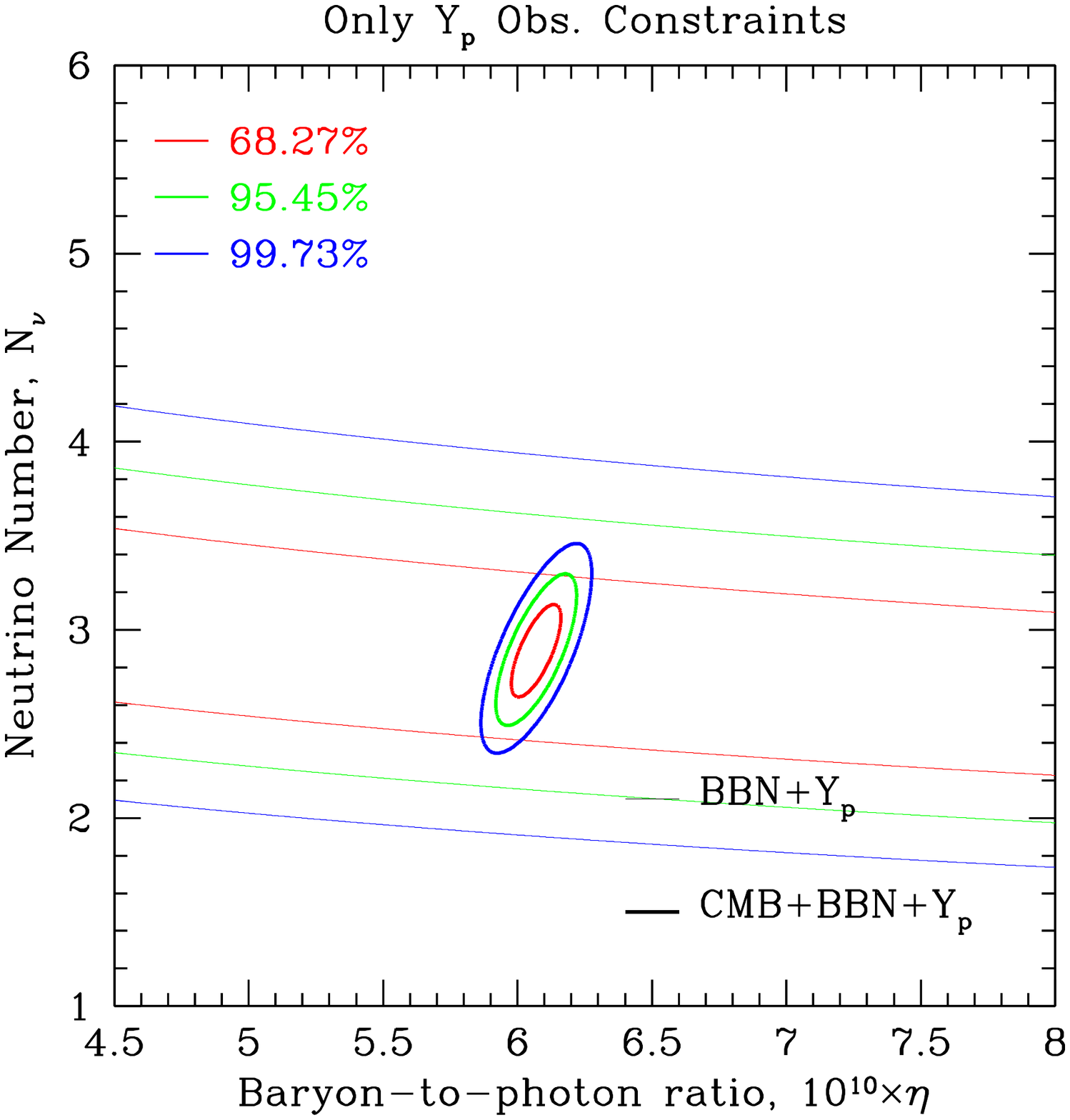,width=0.5\textwidth}
\hskip -.3in
\psfig{file=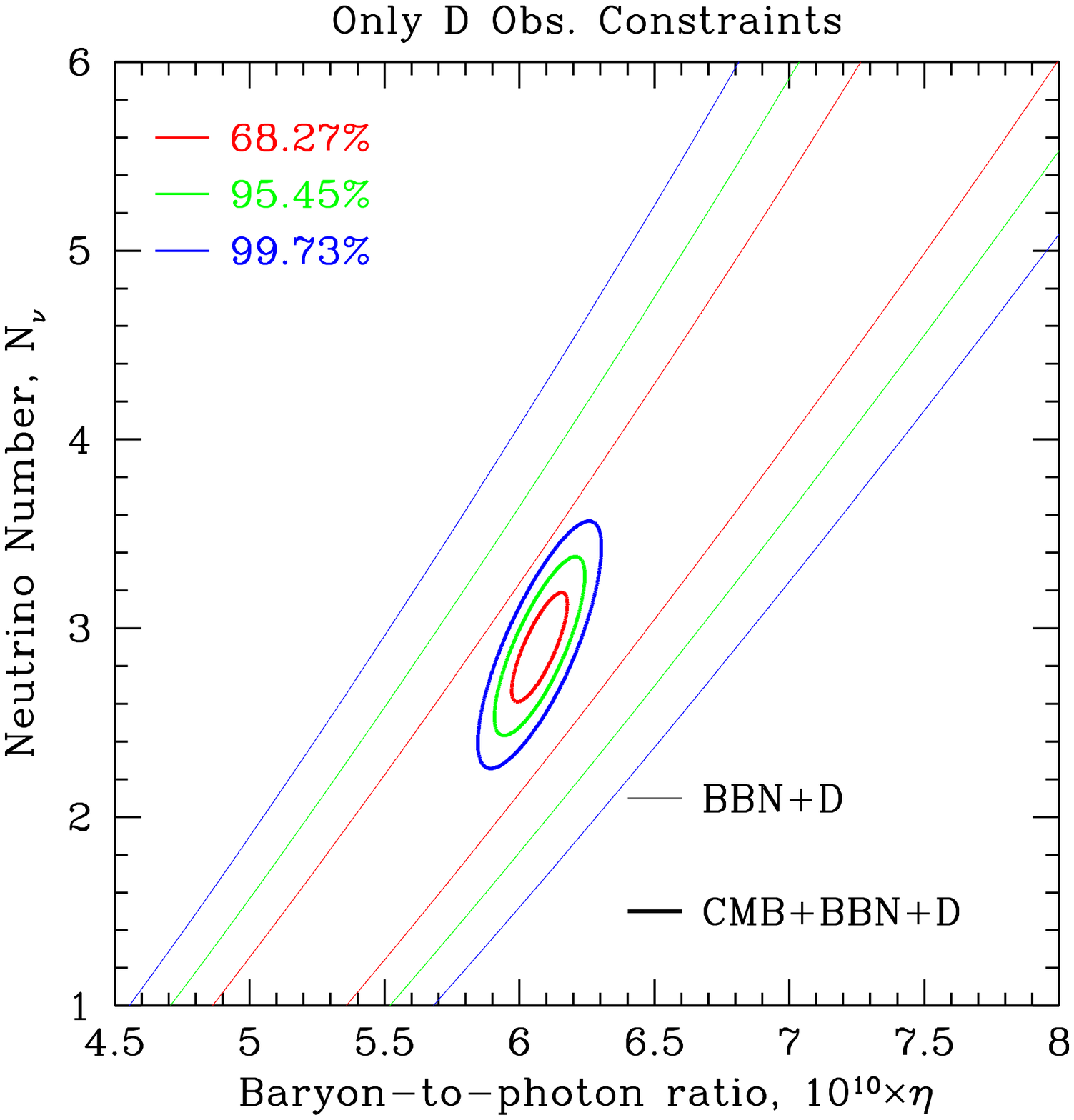,width=0.5\textwidth}\\
\vskip -1in
\psfig{file=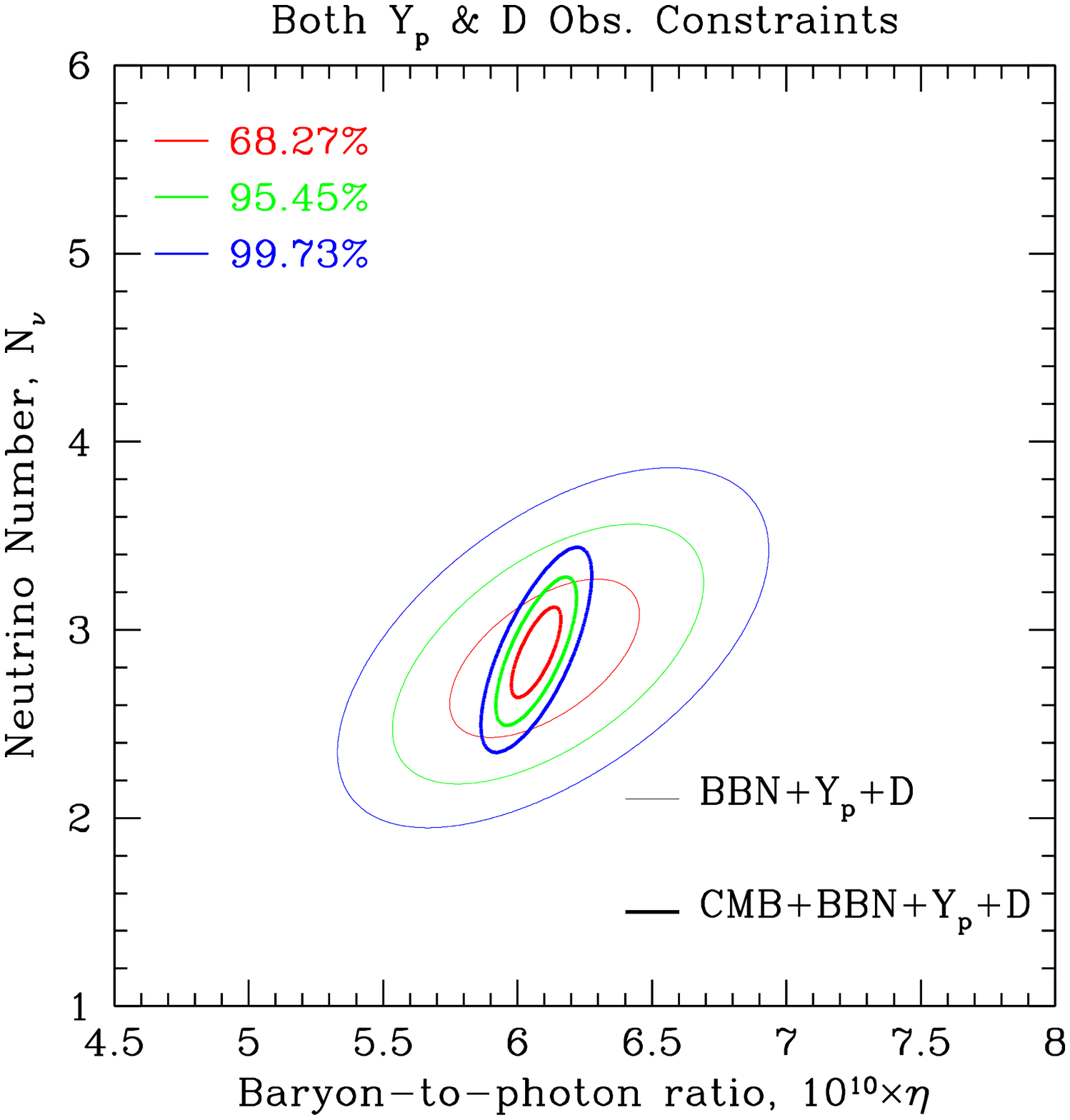,width=0.5\textwidth}
\vskip -1in
\caption{
The resulting 2-dimensional likelihood functions for the baryon to photon ratio ($\eta$) and the number of neutrinos ($N_\nu$), marginalized over the helium mass faction $Y_p$, assuming different combinations of observational constraints on the light elements.
\label{fig:L2D2}
}
\end{figure}
\end{center}

For completeness, we also show 2-d likelihood contours in the $\eta_{10}-N_\nu$ plane in Fig.~\ref{fig:L2D2}. The three panels show the effect of the constraints imposed by the helium and deuterium abundances.  In the first panel, only the helium abundance constraints are applied.  The thinner open curves
are based on BBN alone.  They appear open as the helium abundance alone is a poor baryometer
as has been noted several times already. Without the CMB, the helium abundance data can produce 
an upper limit on $N_\nu$ of about 4 and depends weakly on the value of $\eta$.  When the CMB data
is applied, we obtain the thicker closed contours. The precision determination of $\eta$ from the 
anisotropy spectrum correspondingly produces a very tight limit in $N_\nu$. Here, we see 
clearly that the Standard Model value of $N_\nu = 3$ falls well within the 68\% CL contour.

The next panel of Figure \ref{fig:L2D2} shows the likelihood contours using the deuterium abundance data.
Once again, the thin open curves are based on BBN alone. In this case, they appear open as 
the deuterium abundance is less sensitive to $N_\nu$, though we do note that the contours are not 
vertical and do show some dependence on $N_\nu$ as discussed above. In contrast to \he4, 
for fixed $N_\nu$, the deuterium abundance is capable of fixing $\eta$ relatively precisely. 
Of course when the CMB data is added, the open contours collapse once again into a series
of narrow ellipses. 

The last panel of Figure \ref{fig:L2D2} shows the likelihood contours using both the \he4 and D/H data. 
In this case, even without any CMB input, we are able to obtain reasonably strong constraints on 
both $\eta$ and $N_\nu$ as seen by the thin and larger ellipses. When the CMB data is added
we recover the tight constraints which are qualitatively similar to those in the previous two panels. 

Finally, above in Figure \ref{fig:L2Detanu}, 
we show the 
2-dimensional likelihood function using either CMB only
(the thin outer curves which trace the density of models results of the Monte-Carlo), or 
the combination of CMB and BBN (tighter and thicker curves). In the latter case, no abundance data is used.

\section{Discussion}

Big bang cosmology can be said to have gone full circle. The prediction of the CMB 
was made in the context of the development of BBN and of what became Big Bang Cosmology
\cite{gamow}. Now, the CMB is providing the precision necessary to make accurate 
prediction of the light element abundances in SBBN. In the Standard Model with $N_\nu = 3$,
BBN makes relatively accurate predictions of the light element abundance 
as displayed by the thickness of the bands in Figure \ref{fig:schramm}.
These can be compared directly (or convoluted through a likelihood function)
to the observational determination of the light element abundances. 
The agreement between the theoretical predictions and the abundance D/H is stunning. 
Recent developments in the determination of D/H has produced unparalleled accuracy \cite{cooke}.
This agreement is seen instantly when comparing the likelihood functions of the observations
with that of the predictions of BBN using CMB data as seen in the second panel of Figure \ref{fig:LYi}.
The helium data has also seen considerable progress. New data utilizing a near infrared 
emission line \cite{itg} has led to a marked drop in the uncertainty of the extrapolated primordial
\he4 abundance \cite{aos4}. While the error remains large compared with the precision
of the BBN prediction, the agreement between theory and observation is still impressive.

Is two out of three okay?  Despite the success of the BBN predictions
for \he4 and D/H, there remains a problem with \li7 \cite{cfo5,fieldsliprob}. 
The predicted primordial abundance is about a factor of three higher than the abundance
determined from absorption lines seen in a population of low metallicity halo stars. 
The primordial abundance has since 1981 been associated with a narrow plateau \cite{spites}
of abundance measurements. Recently, the extent of this plateau has been called into question
as a significant amount of downward dispersion is seen at very low metallicity ($[{\rm Fe/H}] < -3$)
\cite{aoki2009,sbordone2010}. If stellar depletion is the explanation of the discrepancy 
between the plateau value and the BBN prediction, it remains to be explained why 
there are virtually no low metallicity stars with abundances above the plateau for all metallicities
below $[{\rm Fe/H}] < -1.5$. If depletion is not the answer, then perhaps the lithium discrepancy points to 
new physics beyond the Standard Model. 

In this review, we have presented the latest combined analysis of BBN predictions
using raw CMB data provided by {\em Planck} \cite{Planck2015,Planckdata}. He have constructed
a series of likelihood function which include various combinations of the CMB, 
the BBN relation between the baryon density and the helium abundance, and various
combinations of \he4 and D/H data. We presented detailed fits and sensitivities of the light element
abundances to the various input parameters as well as the dominant input nuclear rates.
This allowed us to make relatively precise
comparisons between theory and observations in standard BBN. The uncertainty in the prediction 
of \he4 remains dominated by the uncertainty in the neutron mean life. 
We also considered a one-parameter extension of SBBN, allowing the number of 
relativistic degrees of freedom characterized by the number of neutrino flavors to differ from 
the Standard Model value of $N_\nu = 3$. Despite the additional freedom, 
strong constraints on $\eta$ and $N_\nu$ were derived. When all abundance data is used
in conjunction with BBN and CMB data, we obtain a 95\% CL upper limit of $N_\nu < 3.2$. 
As one of the deepest probes in Big Bang Cosmology, BBN continues to thrive. 

Going beyond 2015, we expect further improvements in the data which will better
test the Standard Model. More high resolution data on \he4 emission lines could yield
a further drop in the uncertainty in primordial helium. One should recall that there are still only 
a little over a dozen objects which are well described by models of the emission line regions.
That said, there are less than half a dozen quasar absorption systems which yield high
precision D/H abundances. 
Moreover, the nuclear physics uncertainties in D/H now dominate the error
budget.  Thus there is strong motivation
for future measurements of the rates  most important for deuterium:
$d(p,\gamma)\he3$, as well as $d(d,n)\he3$, $d(d,n)t$, 
and $n(p,\gamma)\he3$ \cite{Nollett2011,DiV}.
We can be hopeful that future measurements lead to 
a reduction in the already small uncertainty in primordial D/H;
futuristically, there is hope 
of detecting cosmological 92 cm deuterium
hyperfine lines that would probe D/H at extremely high redshift
\cite{sig}.
Lastly, we can be certain to expect updated results from the CMB data when the {\em Planck} collaboration
produces its final data release.

\section*{Appendix}
\appendix
\label{app}

 In order to reproduction the likelihoods,
we need order 3 and 4 (skewness and kurtosis) terms in the
multi-dimensional expansion.  

In 1D, this expansion looks like: 
\beq 
{\mathcal L}(x) = \frac{\exp \left\{-\frac{1}{2} z^2\right\}}{\sqrt{2\pi}\sigma} 
\times \left[ 1. + SH_3(z) + KH_4(z) \right], 
\eeq 
where $z=(x-\mu)/\sigma$, $\mu$ is the mean value and $\sigma$ is the
standard deviation and $S$ and $K$ are the skewness and kurtosis
coefficients needed to adequately describe the distribution via a
Hermite polynomial expansion.  The skewness and kurtosis coefficients
are proportional to the Markov Chain average of the respective Hermite
polynomials ($S\propto \avg{H_3(x)}$ and $K\propto \avg{H_4(z)}$).  In
multiple dimensions, the simple gaussian base distribution is replaced
with the fully correlated multi-dimensional gaussian:
\beqar
{\mathcal L}(\vec{x}) &=& \frac{\exp \left\{-\frac{1}{2} (\vec{x}-\vec{\mu})^T {\mathcal C}^{-1} (\vec{x}-\vec{\mu}) \right\}}{\sqrt{(2\pi)^dDet({\mathcal C})}}
\\
\nonumber &\times&\left[ 1. + \sum_{n,m,p=0}^3 S_{nmp,ijk}H_n(z_i)H_m(z_j)H_p(z_k)\delta_{3,n+m+p}\right. \\
\nonumber &+& \left. \sum_{n,m,p,q=0}^4 K_{nmpq,ijkl}H_n(z_i)H_m(z_j)H_p(z_k)H_q(z_l)\delta_{4,n+m+p+q} \right] \, .
\eeqar

\section*{Acknowledgments}
It is a pleasure to thank our recent BBN collaborators Nachiketa Chakrabory,
John Ellis, Doug Friedel, Athol Kemball, Lloyd Knox, Feng Luo,
Marius Millea, Tijana Prodanovi\'c, Vassilis Spanos, and Gary Steigman. 
The work of R.H.C was supported by the National Science Foundation under Grant No. PHY-1430152 (JINA Center for the Evolution of the Elements).
The work of K.A.O.~was supported in part by DOE grant DE-SC0011842  at the University of
Minnesota.
The work of B.D.F.~and T.H.Y.~was partially supported by the U.S. National Science Foundation Grant PHY-1214082.

\end{document}